\documentclass[final,5p,times,twocolumn]{elsarticle}
 \biboptions{comma,sort&compress}
 \usepackage{natbib}
\usepackage{graphicx}
\usepackage{here}
\def\beq{\begin{equation}}
\def\eeq{\end{equation}}
\def\bea{\begin{eqnarray}}
\def\eea{\end{eqnarray}}
\def\bq{\begin{quote}}
\def\eq{\end{quote}}

\def\nnb{\nonumber}
\def\ga{\left(}
\def\dr{\right)}

\def\lrar{\Longrightarrow}
																
\def\nnb{\nonumber}
\def\la{\langle}
\def\ra{\rangle}
\def\nin{\noindent}
\def\ba{\vspace*{-0.2cm}\begin{array}}
\def\ea{\end{array}\vspace*{-0.2cm}}

\def\b{$\bullet~$}
\def\als{\alpha_s}

\def\gg2{ \la\alpha_s G^2 \ra}
\def\gg3{g^3f_{abc}\la G^aG^bG^c \ra}
\def\ggg4{\la\als^2G^4\ra}

\journal{Physics Letters B}

\begin{document}

\begin{frontmatter}

%%
%%%%%%%%%%%%%%%%%%%%%%%%%%%%%%%%%%%%%%%%%%%%%%%%%
%\begin{document}
\title{Gluon Condensates and $\overline{m}_{c,b}$  from  QCD-Moments and their Ratios to  
%${\cal O}
Order $
\alpha_s^3$
{and}
$\la G^4\ra$}

 \author[label1]{Stephan Narison
 %\corref{cor1} 
 }
   \address[label1]{Laboratoire
Univers et Particules de Montpellier, CNRS-IN2P3,  
Case 070, Place Eug\`ene
Bataillon, 34095 - Montpellier Cedex 05, France.}
\ead{snarison@yahoo.fr}

%\author{
% Stephan Narison \thanks{{\it E-mail address:}snarison@yahoo.fr.}
% \address {\footnotesize Laboratoire
%de Physique Th\'eorique et Astroparticules, CNRS-IN2P3 \& Universit\'e
%de Montpellier II, Case 070, Place Eug\`ene
%Bataillon, 34095 - Montpellier Cedex 05,
%France.},
%}
%\begin{document}

\pagestyle{myheadings}
\markright{ }
\begin{abstract}
\noindent
We reconsider the extraction of the gluon condensates $\la \alpha_s G^2\ra$,   $ \la g^3f_{abc} G^3\ra $  and the ${\overline{MS}}$ running  quark masses $\overline{m}_{c,b}$ from different ${\cal M}_n(Q^2)$ Moments and their Ratios by including PT corrections to order $\alpha_s^3$, NPT terms up to $\la G^4\ra$ and using stability criteria of the results versus the degree $n$ (number of $Q^2$-derivative).  We explicitly  show that the spectral part of the lowest  moment ${\cal M}_1(0)$ depends strongly (as expected) on its high-energy (continuum) contribution, which is minimized for ${\cal M}_{n\geq 3-4}(0)$. Using
higher moments and the correlations of $\la \alpha_s G^2\ra$ 
with  $ \la g^3f_{abc} G^3\ra $ and $\la G^4\ra$, we obtain $\la \alpha_s G^2\ra=(7.0\pm 1.3)\times 10^{-2}$ GeV$^4$ and $\la g^3f_{abc} G^3\ra=(8.8\pm 5.5)$ GeV$^2\times\la \alpha_s G^2\ra$, while our analysis favours a modified factorisation for $\la G^4\ra$. 
Using the previous results, we re-determine $\overline{m}_c(\overline {m}_c)$ and find that the commonly used  ${\cal M}_1(0)$ lowest moment tends to overestimate its value  compared to the ones from higher moments where stable values of $\overline {m}_c(\overline {m}_c)$ versus the variations of $n$ and the continuum models are reached. These features can indicate that the quoted errors of $\overline{m}_{c,b}$ from ${\cal M}_1(0)$ may have been underestimated. Our best results from different high-$n$ moments and their ratios are: $\overline{m}_{c}(\overline{m}_{c})=1261(16)~{\rm MeV}$
 and  $\overline{m}_b(\overline{m}_b)=4171(14)$ MeV, in excellent agreement with results obtained in \cite{SNcb} using some judicious choices of ratios of moments.
 \end{abstract}
% \begin{document}
\begin{keyword}  QCD spectral sum rules, gluon condensates, heavy quark masses. 


\end{keyword}

\end{frontmatter}
%%%%%%%%%%%%
%\vspace*{-1.5cm}
\section{Introduction}
\vspace*{-0.25cm}
 \nin
%%%%%%%%%%%%
Non-zero values of the gluon condensates have been advocated by SVZ \cite{SVZ,ZAKA} for non-perturbative QCD. Indeed, the gluon condensates play an important r\^ole in gluodynamics (low-energy theorems,...) and in some bag models as they are directly related to the vacuum energy density (with standard notations):
\beq
E=-{\beta(\alpha_s)\over 8\alpha_s^2}\la \alpha_s G^2\ra~.
\eeq
Moreover, the gluon condensates enter in the OPE of the hadronic correlators \cite{SVZ} and then are important in the analysis of QCD spectral sum rules (QSSR), especially, in the heavy quarks and in the pure Yang-Mills  gluonia/glueball channels where the light quark loops and quark condensates\,$^1$\footnotetext[1]{The heavy quark condensate contribution can be absorbed into the gluon one through the relation \cite{SVZ}: 
$
\la \bar QQ\ra=-{\la\alpha_s G^2\ra/ (12\pi m_Q)}+...
$An analogous relation also occurs for the mixed quark-gluon condensate \cite{SNB1,SNB2,SNB3}.}
 are absent to leading order \cite{SNB1,SNB2,SNB3}. The SVZ value:
\beq
\la\alpha_s G^2\ra\simeq 0.04 ~{\rm GeV}^4~,
\label{eq:standard}
\eeq
extracted (for the first time) from charmonium sum rules \cite{SVZ} has been challenged by different
authors \cite{SNB1,SNB2,SNB3}. Though there are strong indications that the exact value of the gluon condensate is around  (or most likely 2-3 times) this value as obtained from earlier heavy quarks ${\cal M}_n(Q^2)$ \cite{RRY,NIKOL,SHAW}, FESR \cite{YND} and exponential  \cite{BELL} moments, heavy quark mass-splittings \cite{SNHeavy} and $e^+e^-$ \cite{LNT,PEROTTET,BORDES,MENES,SNI} inclusive data.  Most recent determinations 
from $\tau$-decay \cite{ALEPH,OPAL,DAVIER} (see however \cite{SNTAU}) give a value $\la\alpha_s G^2\ra\simeq (0.02 \pm 0.04)~{\rm GeV}^4$, while some particular choices of ${\cal M}_n(Q^2)$  charmonium moments give $(0.04\pm 0.03)~{\rm GeV}^4$  \cite{IOFFE}. Lattice calculations found large range of values \cite{RAKOW,GIACO,GIACO2}. All these results indicate that the value $\la\alpha_s G^2\ra$ is not yet well determined and needs to be reconsidered. \\
In a previous paper \cite{SNcb}, we have extracted, for the first time within QSSR, the correlation between $\la\alpha_s G^2\ra$ and  $ \la g^3f_{abc} G^3\ra $ by working with higher moments known to order $\alpha_s^2$ and up to  $ \la g^3f_{abc} G^3\ra $. We have obtained:
\beq
g^3f_{abc}\la G^3\ra=(31\pm 13)~{\rm GeV^2} \la\alpha_s G^2\ra 
\eeq
or, in terms of the instanton radius:
\beq
 \rho_c\simeq 0.98(21){\rm GeV^{-1}}
 \label{eq:cond1}
\eeq
if one uses the dilute gas instanton  (DGI) model relation~\footnote{Notice that estimates of $\rho_c$ based on DGI give the range of values: 1.5 \cite{SHURYAK}, 2.5 \cite{IOFFE2} and 4.5 GeV$^{-1}$ \cite{SVZ}.}:
\beq
{\la g^3f_{abc}G^3\ra\over \la \alpha_s G^2\ra}= {4\over 5}{12\pi\over\rho_c^2}~.
\label{eq:cond}
\eeq
One may interpret the previous value of  $ \la g^3f_{abc} G^3\ra $ as the one of an {\it effective condensate}
which can absorb into it all higher dimensions condensates not accounted for when the OPE is truncated at the $D=6$-dimension. \\
In the present paper, we shall study the effects of the $D=8$ condensates on the previous results considering the fact that these effects can be sizeable when working with higher moments \cite{NIKOL2,RRY,SHAW}. In the same time, we shall reconsider the determination of  $\la\alpha_s G^2\ra$ and $\overline{m}_{c,b}$ from different ${\cal M}_n(Q^2)$ moments and their ratios by including corrections to order $\alpha_s^3$ and non-perturbative terms up to $\la G^4\ra$. We shall also focus on the extraction of $\overline{m}_{c,b}$ from the widely used ${\cal M}_{n=1}(Q^2=0)$  moments. 
%\end{document}

%%%%%%%%%%%%%%%%%%%%%%%%%%%%%%%%%%%%%%%
%\vspace*{-0.5cm}
%\vspace*{-0.3cm}
\section{Moment sum rules, stability criteria and optimal results}
\label{sec:qssr}
\vspace*{-0.25cm}
%%%%%%%%%%%%%%%%%%%%%%%%%%%%%%%%%%%%%%%
 \nin
\b Here, we shall be concerned with the two-point correlator of a heavy quark $Q\equiv c,b$:
 \bea
&& -\ga g^{\mu\nu}q^2-q^\mu q^\nu\dr \Pi_Q(q^2)\equiv \nnb\\
&& i\int d^4x ~e^{\rm-iqx}\la 0\vert {\cal T} J^\mu_Q(x)\ga J^\nu_Q(0)\dr^\dagger \vert 0\ra~,
 \eea
 where : $J_Q^\mu=\bar Q \gamma^\mu Q$ is the heavy quark neutral vector current. 
 %For $Q\equiv c$, Im $\Pi(s)$ can be related to the charmonium leptonic widths and masses. 
 %%%%%%%%%%%%%%%%%%%%%%%%%%%%%%%%%
 Different forms of QSSR exist in the literature \cite{SNB1,SNB2,SNB3}. In a previous \cite{SNcb} and in the present paper, we work with the moments~\footnote{We shall use the same normalization as  \cite{IOFFE}.}:
 \bea
 {\cal M}_n\ga -q^2\equiv Q^2\dr&\equiv& 4\pi^2{(-1)^n\over n!}\ga {d\over dQ^2}\dr^n \Pi(-Q^2)\nnb\\
 &=&\int_{4m_Q^2}^\infty dt {{R}(t,m_c^2)\over (t+Q^2)^{n+1}}~,
 \eea
and with their ratios:
 \beq
 r_{n/n+1}(Q^2)={{\cal M}_n(Q^2)\over {\cal M}_{n+1}(Q^2)},~~~~r_{n/n+2}(Q^2)={{\cal M}_n(Q^2)\over {\cal M}_{n+2}(Q^2)}~,
 \eeq
where the experimental sides are more precise than the absolute moments ${\cal M}_n(Q^2)$. \\
\b In the following, we shall use stability criteria, i.e. a minimum dependence of the results
on the variation of the finite number of derivatives $n$. In practice, this minimum sensitivity is signaled by
the presence of a a plateau or  a minimum. \\
\b We shall study later the effect of the QCD continuum models on the results.\\
%will give small contribution such that the dependence of the result on the variation of the continuum threshold $t_c$ will be negligible. \\
\b We shall denote by optimal result the one obtained within the previous stability criteria and
which is less affected by the different forms of the continuum models.

%Also, in the ratios, partial cancellations of different perturbative, continuum contributions as well as non-perturbative terms occur, which render the QCD approximation more precise than in the absolute moments.
%%%%%%%%%%%%%%%%%%%%%%%%
 %%%%%%%%%%%%%%%%%%%%%%%%%%%%%%%%%%%%%%%
%\vspace*{-0.5cm}
%\vspace*{-0.3cm}
\section{QCD expressions of the sum rules }
\label{sec:qcd}
\vspace*{-0.25cm}
%%%%%%%%%%%%%%%%%%%%%%%%%%%%%%%%%%%%%%%
 \nin
The QCD expressions of the moments can be derived from the ones of the vector spectral function $R$.\\
\b To  lowest order, it reads :
\beq
R_{LO} ={ v\over 2}(3 - v^2)
\label{eq:velocity}
\eeq
where $
v \equiv \sqrt{1 - 4m_Q^2/t}$ is the quark velocity. \\
\b The $\alpha_s$ correction is known exactly to ${\cal O}(\alpha_s)$ \cite{KALLEN} and an interpolating formula has been proposed in \cite{SCHWINGER}. \\
\b To order $\alpha_s^2$, we shall use the approximate formula given in \cite{CHET3} and derived from the exact expression in \cite{CHET0,CHET1,CHET2}.  \\
\b To order $\alpha_s^3$, the three lowest ${\cal M}_1(0)$   \cite{KUHN2} and ${\cal M}_{2,3}(0)$ moments \cite{MAIER} are known analytically . Semi-analytic expressions of higher moments ${\cal M}_n(0)$ using Pad\' e approximants \cite{MATEU} and Mellin-Barnes transform \cite{GREYNAT} are also available. \\
\b The gluon condensate  $\la \alpha_s G^2\ra$ contribution to the two-point correlator is known to lowest order \cite{SVZ} and to order $\alpha_s$~\cite{BROAD}. \\ 
\b The dimension-six condensates ($\la g^3f_{abc}G^3\ra$ and $g^4 \la\bar u u\ra^2$) contributions have been obtained by \cite{NIKOL}. Convenient expressions for numerical analysis of different ${\cal M}_n(Q^2)$  moments including the $\la g^3f_{abc}G^3\ra$ term are given by \cite{IOFFE}. We have checked some but not all of them. \\
\b The $\la G^4\ra$ condensate
 contributions have been calculated by \cite{NIKOL2} to lowest order. The expressions of ${\cal M}_n(Q^2)$   have been given by \cite{NIKOL2} and \cite{RRY}. \\
In the following discussions, we shall not transcript all these previous long and tedious formulae  which interested readers can found in the original papers. 
%%%%%%%%%%%%%%%%%%%%%%%%%%%%%%%%%%%%%%%

 %%%%%%%%%%%%%%%%%%%%%%%%%%%%%%%%%%%%%%%
%\vspace*{-0.5cm}
%\vspace*{-0.3cm}
\section{Experimental parametrization of the sum rules }
\label{sec:exp}
\vspace*{-0.25cm}
%%%%%%%%%%%%%%%%%%%%%%%%%%%%%%%%%%%%%%%
 \nin 
 In a narrow width approximation (NWA) and for $Q\equiv c$~\footnote{A missprint of factor $\pi$ is in \cite{SNcb} but does not affect
 the results.}:
 \bea
 { R}(t)&\equiv& 4\pi{\rm Im} \Pi(t+i\epsilon)\nnb\\
 &=&\pi {N\over Q_c^2 \alpha^2}\sum_{J/\psi}M_{\psi}\Gamma_{\psi\to e^+e^-}\delta(\ga t-M^2_{\psi}\dr~,
 \eea
 where $N=3$ is the colour number; $M_{\psi}$ and $\Gamma_{\psi\to e^+e^-}$ are the mass and leptonic width of the $J/\psi$ mesons; $Q_c=2/3$ is the charm electric charge in units of $e$; $\alpha=1/133.6$ is the running electromagnetic coupling evaluated at $M^2_{\psi}$. We shall use the experimental values of the $J/\psi$ parameters compiled in Table \ref{tab:psi}.\\
  %%%%%%%%%%%%%%%%%%%%%%%%%%%%%%%
 \vspace*{-.5cm}
{\scriptsize
\begin{table}[hbt]
\setlength{\tabcolsep}{1.5pc}
 \caption{\scriptsize    Masses and electronic widths of the  $J/\psi$ family from PDG10 \cite{PDG}. }
{\small
\begin{tabular}{lll}
&\\
\hline
%\hline
%\\
Name&Mass [MeV]&$\Gamma_{J/\psi\to e^+e^-}$ [keV] \\
%\\
\hline
%\hline
\\
$J/\psi(1S)$&3096.916(11)&5.55(14)\\
$\psi(2S)$&3686.093(34)&2.33(7)\\
$\psi(3770)$&3775.2(1.7)&0.259(16)\\
$\psi(4040)$&4039(1)&0.86(7)\\
$\psi(4160)$&4153(3)&0.83(7)\\
$\psi(4415)$&4421(4)&0.58(7)\\
\\
\hline
\end{tabular}
}
\label{tab:psi}
\end{table}
}
%\vspace*{-2.5cm}
\nin
\\
We shall parametrize the contributions from $\sqrt{t_c}\geq (4.6\pm 0.1)$ GeV using either:\\
\b {\it Model 1:} The approximate PT QCD expression of  the spectral function to order $\alpha_s^2$ up to order $(m_c^2/t)^6$ given in \cite{CHET3} and the $\alpha_s^3$ contribution from non-singlet contribution up to order $(m_c^2/t)^2$ given in \cite{HOANG}. \\
\b {\it Model 2:} The asymptotic PT expression of the  spectral function known  to order $\alpha_s^3$ where the quark mass corrections are neglected~\footnote{Original papers are given in Refs. 317 to 321 of the book in Ref.~\cite{SNB1}.}.\\
\b {\it Model 3:} Fits of different data above the $\psi(2S)$ mass: the most recent fit is done in \cite{HOANG} where a comparison of results from different fitting procedures can be found. 
 %%%%%%%%%%%%%%%%%%%%%%%%%%%%%%%%%%%%%%%
%\vspace*{-0.5cm}
%\vspace*{-0.3cm}
\section{Test of the continuum model-dependence of the  moments }
\label{sec:cont}
\vspace*{-0.25cm}
%%%%%%%%%%%%%%%%%%%%%%%%%%%%%%%%%%%%%%%
 \nin 
 %%%%%%%%%%%%%%%%%%%%%%%%%%%%%%%
\begin{figure}[hbt]
\begin{center}
\includegraphics[width=6.5cm]{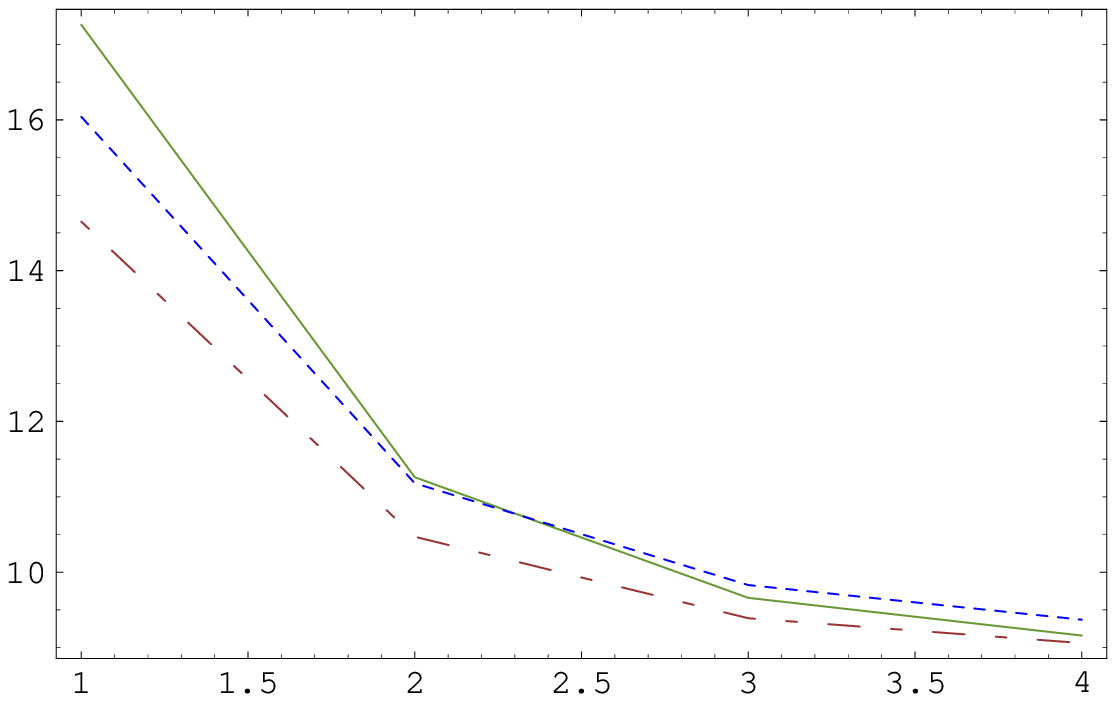}
\includegraphics[width=6.5cm]{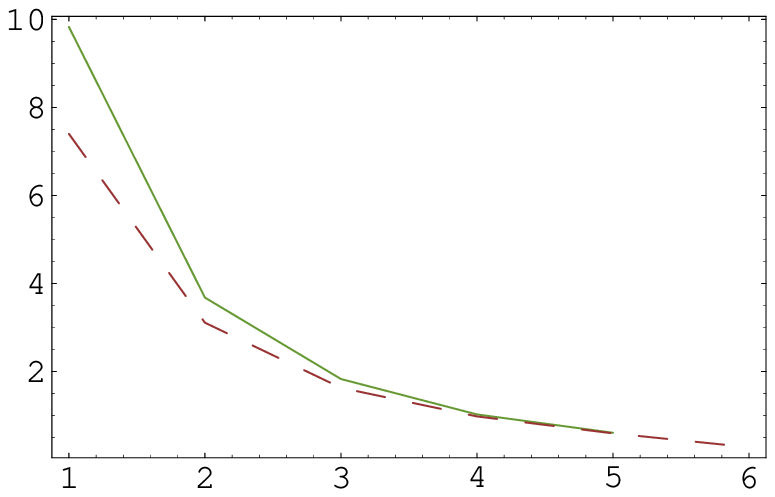}
%\vspace*{-0.7cm}
\caption{\footnotesize  Behaviour of moments ${\cal M}_n(Q^2)$ in units of $(4\overline{m}_c^2)^n\times10^{n+1}$ versus $n$ for different models of the continuum as defined in section \ref{sec:exp}:  {\bf a)}  ${\cal M}_n(0)$ : Model 1: green (continuous), Model 2 : red (dot-dashed), Model 3: bleu (dot); {\bf b)} the same as Fig \ref{fig:cont}a) but for ${\cal M}_n(4\overline{m}_c^2)$ for Models 1 and 2.}
\label{fig:cont}
\end{center}
%\vspace*{-1.cm}
\end{figure} 
\nin
%\\
 %%%%%%%%%%%%%%%%%%%%%%%%%%%%%%%
In this section, we test the model-dependence of the experimental side of the moments using the previous models for parametrizing the continuum (high-energy) contribution  to the
spectral function. The analysis is shown in Fig. \ref{fig:cont}a for the moments ${\cal M}_n(0)$ using Models 1, 2 and 3 for different values of $n$ and  in Fig \ref{fig:cont}b for the moments ${\cal M}_n(4\overline{m}_c^2)$ using Models 1 and 2. One can deduce 
that this model dependence can be avoided when working with values of $n\geq 3,4$. One can also notice that for ${\cal M}_1(0)$, the continuum (high-energy) contribution to the moments is about (40-50)\% of the total contribution, which indicates
that the resulting value of $m_c$ from the low moments $n\leq (2-3)$ will depend strongly on the appreciation of this high-energy
 behaviour which is not measured accurately as also emphasized in \cite{HOANG}.
  %%%%%%%%%%%%%%%%%%%%%%%%%%%%%%%%%%%%%%%
%\vspace*{-0.5cm}
%\vspace*{-0.3cm}
\section{QCD inputs and higher gluon condensates}
\label{sec:g4}
\vspace*{-0.25cm}
%%%%%%%%%%%%%%%%%%%%%%%%%%%%%%%%%%%%%%%
 \nin 
\b From the different expressions of the PT series given in \cite{IOFFE},  we observe that, unlike ${\cal M}_n(Q^2=0)$ where the coefficients increase approximately like $n$ for large $n$ (the same feature occurs for the $\alpha_s^3$ term given in \cite{MAIER,MATEU,GREYNAT}),  the ones of  ${\cal M}_n(Q^2\not=0)$ remains (within a factor 2) almost constant  though change sign from low to high moments. 
Therefore, we estimate the coefficient of the ${\cal O}(\alpha_s^3)$ term of the moments ${\cal M}_n(Q^2\not=0)$ to be about:
\beq
c_3\vert_{Q^2\not=0,n}\simeq \pm c_3\vert_{Q^2=0,n=1}\simeq \pm 5.6~,
\label{eq:as3}
\eeq
which is larger than the estimate used in \cite{SNcb}, where it has been assumed that the ratio of the $\alpha_s^2$  over the $\alpha_s^3$ coefficients are approximatively the same for each moment. \\
\b We shall use the input values \cite{SNcb,SNTAU,SNI}:
\bea
\overline{m}_c (\overline{m}_c)&=&1261(18)~{\rm  MeV}~,\nnb\\
\alpha_s(m_c)\vert_{n_f=4}&=&0.408(14)~{\rm from~\tau-decays}~,\nnb\\
\alpha_s\la\bar \psi\psi\ra^2&=& 4.5\times 10^{-4}~{\rm GeV}^6~{\rm from ~e^+e^-}~.
\label{eq:pt-param}
\label{eq:alphas}
\eea
% obtained respectively in \cite{SNcb},   \cite{SNTAU} as deduced from $\tau$-decays analysis and in \cite{SNI}  from a fit of $e^+e^- \to$ I=1 Hadrons data. 
The error in the value of $\alpha_s$ is the distance between its value and the world average \cite{PDG,BETHKE}. 
 \\
% which would correspond to 1/2 of the factorization value used in \cite{NIKOL2}. \\
\b The QCD expressions of the moments are tabulated in \cite{IOFFE} for the  fixed order PT series up to order $\alpha_s^2$ and including the  $ \la g^3f_{abc} G^3\ra $ condensate. The contribution of the $\alpha_s^2\la\bar \psi\psi\ra^2$ D=6 condensate is numerically negligible and has been omitted. \\
\b The contribution of the D=8 condensates can be found in \cite{NIKOL2} and \cite{RRY}. 
 In general, one can form eight operators for the D=8 gluon condensates:
\bea
O_1&=&\la \mbox{\bf Tr}~G^2~\mbox{\bf Tr}~G^2\ra~,\nnb \\
O_2&=&\la \mbox{\bf Tr}~G_{\nu\mu}G^{\rho\mu}~
\mbox{\bf Tr}~G_{\nu\tau}G^{\rho\tau}\ra~,\nnb\\
O_3&=&\la \mbox{\bf Tr}~G_{\nu\mu}G^{\tau\rho}~
\mbox{\bf Tr}~G_{\nu\mu}G^{\tau\rho}\ra~,\nnb\\
O_4&=&\la \mbox{\bf Tr}~G_{\nu\mu}G^{\tau\rho}~
\mbox{\bf Tr}~G^{\nu}_{\tau}G^{\mu}_{\rho}\ra~,\nnb\\
O_5&=&\la \mbox{\bf Tr}~G_{\nu\mu}G^{\mu\rho}~
G_{\rho\tau}G^{\tau\nu}\ra~,\nnb\\
O_6&=&\la \mbox{\bf Tr}~G_{\nu\mu}G^{\nu\mu}G^{\tau\rho}G_{\tau\rho}~
\ra~,\nnb\\
O_7&=&\la \mbox{\bf Tr}~G_{\nu\mu}G^{\nu\rho}~
G_{\mu\tau}G^{\rho\tau}\ra~,\nnb\\
O_8&=&\la \mbox{\bf Tr}~G_{\nu\mu}G^{\rho\tau}~
G^{\nu\mu}G^{\rho\tau}\ra~.
\eea
Using the symmetry properties of the colour indices and an explicit
evaluation of the trace, one can show that one has only six independent
operators and the relation for $N=3$ colours \cite{BAGAN}:
\bea
O_5+2O_7&=&O_2+\frac{1}{2}O_4~,\nnb \\
O_8+2O_6&=&O_3+\frac{1}{2}O_1~.
\label{constraint}
\eea
Normalized to ${\la G^2\ra^2}$, the use of the vacuum saturation in the large $N$-limit gives:
\bea
&O_1=\frac{1}{4}\ga 1+\frac{1}{3}\frac{1}{N^2-1}\dr,&
O_2=\frac{1}{4}\ga \frac{1}{4}+
\frac{1}{3}\frac{1}{N^2-1}\dr, \nnb\\
&O_3=\frac{1}{4}\ga \frac{1}{6}+
\frac{7}{6}\frac{1}{N^2-1}\dr,&
O_4=\frac{1}{4}\ga \frac{1}{12}+
\frac{1}{2}\frac{1}{N^2-1}\dr,\nnb\\
&O_5=\frac{1}{4N}\ga \frac{1}{2}-
\frac{1}{12}\frac{1}{N^2-1}\dr,&
O_6=\frac{1}{4N}\ga \frac{7}{6}-
\frac{1}{6}\frac{1}{N^2-1}\dr,\nnb\\
&O_7=\frac{1}{4N}\ga \frac{1}{3}-
\frac{1}{4}\frac{2}{N^2-1}\dr,&
O_8=\frac{1}{4N}\ga \frac{1}{3}-
\frac{1}{N^2-1}\dr,
\eea
which indicates that
only the first four operators are leading in $1/N$, and the
previous constraints in Eq. (\ref{constraint}) are not satisfied for large $N$. Moreover, the $1/N^2$
corrections to these leading-term are also large for $N=3$ in the
case of $O_3$ and
$O_4$, and raise some doubts on the validity of the $1/N$-approximation.  
Therefore, a modified factorization has been proposed in \cite{BAGAN}, where the D=8 gluon condensates have been expressed in terms of $O_2$ which is not constrained. Normalized to $\la G^2\ra^2$, one has:
\bea
&O_1=3O_6={1\over 4}&O_3=2O_4=-{1\over 16}+2O_2\nnb\\
&O_5=O_7=-{1\over 192}+{1\over 2}O_2&O_8=-{5\over 48}+2O_2.
\eea
Ref. \cite{BAGAN}  estimates $O_2$ using either its large $N$ or its factorization value. Noting that the dominant contribution to the sum rule is due to $O_5$, Ref. \cite{BAGAN} notices that
the factorization proposed in \cite{NIKOL2} overestimates the D=8 gluon condensate contributions.\\
\b For definiteness, we use the following notations and values: 
\bea
\rho_c&\equiv& {\rm  instanton ~radius ~introduced~ in~ Eq.} (\ref{eq:cond})~,\nnb\\
{\rm fac}&=& 1\equiv {\rm factorisation~ of~  \la G^4 \ra  }~,\nnb\\
{\rm fac}&\simeq& 0.5\equiv {\rm modified ~factorisation~ of}~ \la G^4\ra ~.
\label{eq:fac}
\eea
respectively from  \cite{NIKOL2} and \cite{BAGAN}.
\b  We also use the 
value of the scale $M^2\approx 0.3$ GeV$^2$ estimated in \cite{NIKOL2}, which characterizes the average virtual momentum of the vacuum gluons and quarks and
 which relates, using factorization, some of the $D=8$ to the $D=6$ condensates:
\bea
&\la g^4 j_\mu^a D_\alpha D_\alpha j_{\mu}^a\ra=\la g^4j_{\mu}^aj_{\mu}^a\ra M^2=-{4\over 3} g^4\la \bar uu\ra^2M^2,\nnb\\
&\la g^5f^{abc}G^a_{\mu\nu}j_\mu^b j_\nu^c\ra=-{3\over 2} g^4\la \bar uu \ra^2 M^2,\nnb\\
&\la g^3f^{abc}G^a_{\mu\nu}G^b_{\nu\lambda} D_\alpha D_\alpha G^c_{\lambda\mu}\ra=\la g^3f^{abc}G^a_{\mu\nu}G^b_{\nu\lambda} G^c_{\lambda\mu}\ra M^2,
\eea
where $j_\mu^a=\sum_{u,d,s}\bar\psi\gamma_\mu(\lambda^a/2)\psi$ and $D_\alpha$ the covariant derivative.
 %%%%%%%%%%%%%%%%%%%%%%%%%%%%%%%%%%%%%%%
%\vspace*{-0.5cm}
%\vspace*{-0.3cm}
\section{Hunting $\la\alpha_s G^2\ra$  from  higher  moments ${\cal M}_n(Q^2)$}
\label{sec:g2}
\vspace*{-0.25cm}
%%%%%%%%%%%%%%%%%%%%%%%%%%%%%%%%%%%%%%%
 \nin 
 As mentioned  in the introduction, the gluon condensate plays a key r\^ole in QCD gluon dynamics  like is the quark condensate $\la \bar\psi\psi\ra$ for chiral symmetry breaking. We have also mentioned 
that there is a spread of predictions of its value in the literature. The 
 extraction of $\la\alpha_s G^2\ra$ in this paper is closed to the one using charmonium sum rules in the early literatures which follow the pioneer work of SVZ \cite{SVZ}. \\
\b In our analysis, we shall work with higher moments which are more sensitive to $\la\alpha_s G^2\ra$ but limit ourselves to the ones where the higher dimension-six and -eight condensates remain still small corrections such that  the OPE remains valid. This compromise choice eliminates the use of higher $Q^2=0$ moments where their convergence has been the subject of hot debate in the past \cite{NIKOL,NIKOL2,NOVIKOV}. Instead the $Q^2\not=0$ moments converge faster \cite{RRY} which allow to work with higher $n$-values. In the following, we shall work with ${\cal M}_n(Q^2=0)$ for $n\leq 5$, ${\cal M}_n(Q^2=4\overline{m}_c^2)$ for $n\leq 11-12$ and with ${\cal M}_n(Q^2=8\overline{m}_c^2)$ for $n\leq 20$ where the OPE still makes sense when using the values of the vacuum condensates given in the literature~\cite{SNB1}. \\
\b We extract $\la\alpha_s G^2\ra$ using its correlations with the $D=6$ and 8 condensates introduced above.  We allow the instanton radius $\rho_c$ which correlates  $\la\alpha_s G^2\ra$ and $\la g^3f_{abc} G^3\ra$ to move from 1 to 5 GeV$^{-1}$ where the latter would be the value given by a dilute gas instanton model estimate \cite{SVZ}. We shall also use the relation of $\la\alpha_s G^2\ra$ and $\la g^3f_{abc} G^3\ra$ with the D=8 condensates if one assumes a factorization hypothesis \cite{NIKOL2} or its modified form \cite{BAGAN}. \\
\b Notice that, unlike \cite{IOFFE}, we
 fix $\overline{m}_c$, which is, at present, known with good accuracy,  in order to give stronger constraints on the value of  $\la\alpha_s G^2\ra$.  
We show the results as function of the number $n$ of derivatives for $Q^2=4\overline{m}_c^2$ and $8\overline{m}_c^2$ and for different values of the QCD input parameters.  %\end{document}
%%%%%%%%%%%%%%%%%%%%%%%%%%%%%%%
\begin{figure}[hbt]
\begin{center}
\includegraphics[width=7cm]{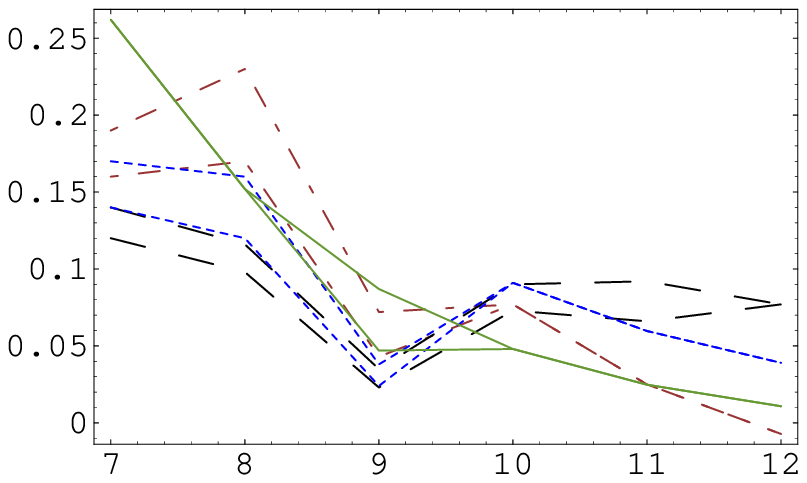}
\includegraphics[width=7cm]{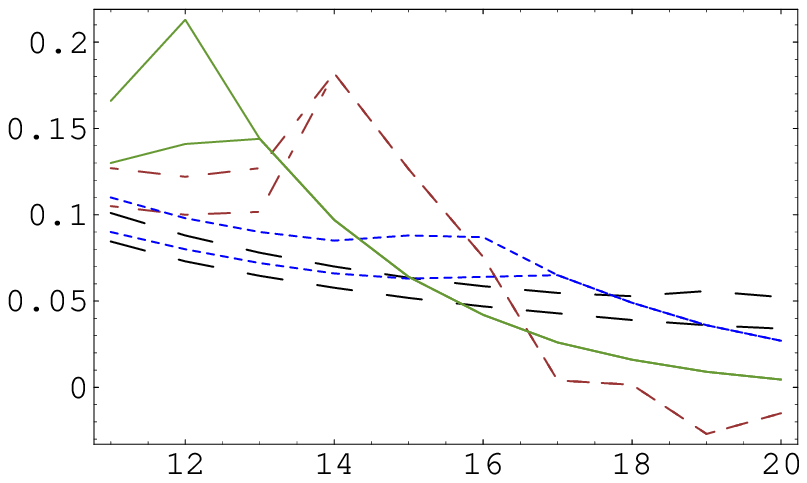}
%\vspace*{-0.7cm}
\caption{\footnotesize  Behaviour of $\la\alpha_s G^2\ra$ in units of GeV$^4$ versus $n$ from ${\cal M}_n(Q^2)$ moments   for different values of the QCD inputs. We use the central value of $\alpha_s$ given in Eq. (\ref{eq:alphas}). The region between the same curves correspond to the values of $c_3$ from Eq. (\ref{eq:as3}):  {\bf a)}  ${\cal M}_n(4m_c^2)$ : $\rho_c=1$ GeV$^{-1}$, fac=1: green (continuous); $\rho_c=5$ GeV$^{-1}$, fac=1:  blue (dot);  $\rho_c=1$ GeV$^{-1}$, fac=0.5: red  (dot-dashed); $\rho_c=5$ GeV$^{-1}$, fac=0.5:  black (dashed);
{\bf b)} The same as Fig \ref{fig:ag2}a) but for ${\cal M}_n(8m_c^2)$}
\label{fig:ag2}
\end{center}
%\vspace*{-1.cm}
\end{figure} 
\nin
\\
 %%%%%%%%%%%%%%%%%%%%%%%%%%%%%%%
One can notice from the Fig \ref{fig:ag2} that the effect of $\alpha_s^3$ is relatively small. Much more stable values of $\la\alpha_s G^2\ra$ correspond to the case of a modified factorisation of $\la G^4\ra$ which sounds better founded from the analysis of \cite{BAGAN} based on the $1/N$ approach. Taking into account these remarks, we deduce in units of GeV$^4$:
\bea
 \la\alpha_s G^2\ra&=&(4.8-9.2)\times 10^{-2}~~~{\rm from} ~{\cal M}_n(4m_c^2) ,\nnb\\
&&(5.6-8.3)\times 10^{-2}~~~{\rm from} ~{\cal M}_n(8m_c^2)~,
\label{eq:ag2range}
\eea
where we have used the {\it Mathematica Package FindRoots}, which we shall also use later on for deriving all the results in this paper.\\
\b We consider as a final result the most precise determination from ${\cal M}_n(8m_c^2)$ which can be written as:
\beq
\la\alpha_s G^2\ra=(7.0\pm 1.3)\times 10^{-2}~{\rm GeV}^4~.
\label{eq:ag2}
\eeq
This result goes in line with different claims \cite{NIKOL,SHAW,BELL,BERT,NEUF,SNHeavy,PEROTTET,SNI,SNB1,SNB2} that the SVZ value given in Eq. (\ref{eq:standard}) understimates the value of the gluon condensate~\footnote{A compilation of different determinations can be found in Table 2 of \cite{SNHeavy} and in the book \cite{SNB1} (reprinted papers in Chapters 51 and 52).}. This result agrees quite well with
the one derived from the charmonium and bottomium mass-splittings using double ratio of sum rules (DRSR) \cite{SNHeavy}:
\beq
\la\alpha_s G^2\ra=(7.5\pm 2.5)\times 10^{-2}~{\rm GeV}^4~,
\eeq
and from $\tau$-like sum rule for $e^+e^-\to I=1$ hadrons data \cite{SNI}:
 \beq
\la\alpha_s G^2\ra= 6.1(0.7)\times 10^{-2} ~{\rm GeV}^4~.
\eeq
Our result is more precise than the one in \cite{IOFFE}, using some particular choice of moments, as, here, we have fixed  the value of $\overline{m}_c$ while in \cite{IOFFE} a two-parameter fit $(\overline{m}_c, \la\alpha_s G^2\ra)$ has been performed. Indeed, using as input the value of $\overline{m}_c(\overline{m}_c)$ in Eq. (\ref{eq:pt-param}), one would deduce from the different figures given in \cite{IOFFE}:
\beq
\la\alpha_s G^2\ra\simeq (3.5-7.5)\times 10^{-2}~{\rm GeV}^4~,
\eeq
obtained to order $\alpha_s^2$ and without the inclusion of $\la G^4\ra$. This range of values
is in agreement with the one in Eq. (\ref{eq:ag2range}) but less precise. 
 %%%%%%%%%%%%%%%%%%%%%%%%%%%%%%%%%%%%%%%
%\vspace*{-0.5cm}
%\vspace*{-0.3cm}
\section{Re-extraction of   $ \la g^3f_{abc} G^3\ra $  and factorisation test of $\la G^4\ra$}
\label{sec:g3}
\vspace*{-0.25cm}
%%%%%%%%%%%%%%%%%%%%%%%%%%%%%%%%%%%%%%%
%%%%%%%%%%%%%%%%%%%%%%%%%%%%%%%
\begin{figure}[hbt]
\begin{center}
\includegraphics[width=7cm]{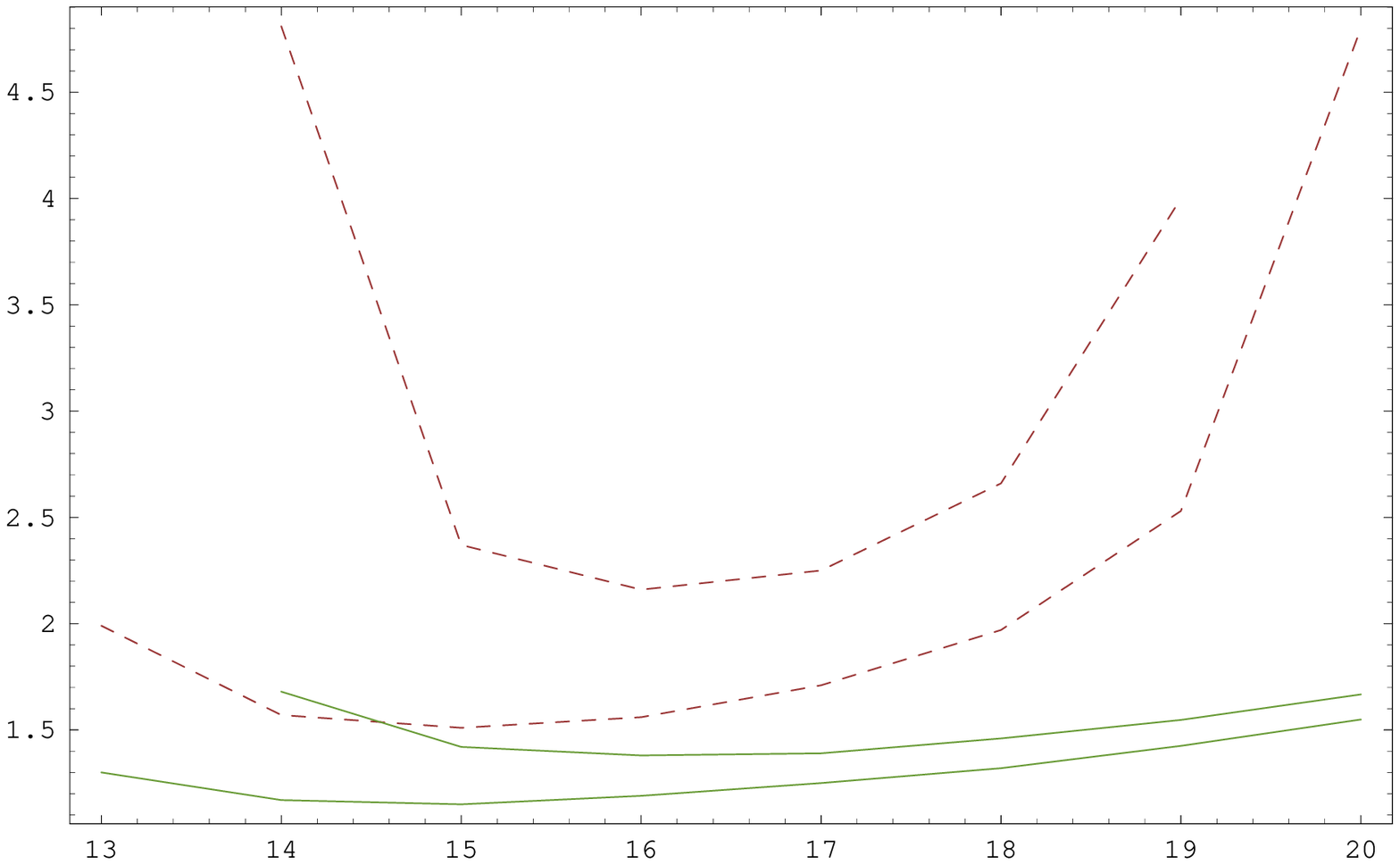}
%\vspace*{-0.7cm}
\caption{\footnotesize  Behaviour of the instanton radius $\rho_c$ in units of GeV$^{-1}$ versus $n$ from ${\cal M}_n(8m_c^2)$ moments for the central values of $\overline{m}_c(\overline{m}_c)$ and $\alpha_s$ given in Eq. (\ref{eq:pt-param}).
The curves correspond to fac=0.5 for the factorisation of $\la G^4\ra$ and $\la\alpha_s G^2\ra=0.070$ GeV$^4$. 
The region between the same curves correspond to the values of $c_3$ from Eq. (\ref{eq:as3}) : dashed (red) with $\la G^4\ra$ and green (continuous) without $\la G^4\ra$.}
\label{fig:rho}
\end{center}
%\vspace*{-1.cm}
\end{figure} 
\nin
 %%%%%%%%%%%%%%%%%%%%%%%%%%%%%%%
Using the previous new informations, we re-extract the value of  $ \la g^3f_{abc} G^3\ra $ firstly obtained
in \cite{SNcb} using the sum rules approach. As remarked in \cite{SNcb}, the moment ${\cal M}_n(8m_c^2)$ can provide the most accurate value of  $ \la g^3f_{abc} G^3\ra $. We plot in Fig \ref{fig:rho}
the value of the instanton radius $\rho_c$ defined in Eq. (\ref{eq:cond}) versus the number of moments for given values of $\overline{m}_c$,  $\alpha_s$, $\la\alpha_s G^2\ra$ and 
the factorisation factor fac of the $\la G^4\ra$ condensates, . We have only shown in Fig \ref{fig:rho}, the curves for fac=0.5 because the one for fac =1 gives unrealistic values of $\rho_c$. This can be an indirect indication that the value fac=1 is less favoured, a result  which supports the $1/N$ result in \cite{BAGAN}. A similar feature is also signaled when extracting $\la\alpha_s G^2\ra$ because  for fac=0.5, larger stabilities versus the change of $n$ (see Fig \ref{fig:ag2}) are obtained. At the minimas of the curves in Fig \ref{fig:rho}, we deduce the optimal value of $\rho_c$ in GeV$^{-1}$ when the effect of $\la G^4\ra$ is included:
\beq
\rho_c=1.84\pm 0.24_{\alpha_s}\pm0.33_{\alpha_s^3}\pm 0.27_{G^2}~,
\eeq
which after adding the errors quadratically gives:
\bea
\rho_c&=&(1.84\pm 0.49)~{\rm GeV}^{-1} \nnb\\
&\lrar& {\la g^3f_{abc} G^3\ra\over  \la\alpha_s G^2\ra}=(8.8\pm 4.7)~{\rm GeV^2}~.
\label{eq:rhoc}
\eea
We consider this result as improvement of the previous result quoted in Eq. (\ref{eq:cond1}), which has been affected by the presence of $\la G^4\ra$ in the OPE (see the two continuous (green) curves in Fig \ref{fig:rho} when $\la G^4\ra$ is not included \footnote{These values agree with the one obtained in \cite{SNcb} using some judicious choice of the ratios of moments $r_{13/14} $ and  $r_{14/15} $.})
%However, we have checked that the results are not stable for other choices and we shall not consider here. ) 
as (a priori) expected. 
This value of  $ \la g^3f_{abc} G^3\ra $ is in the range of lattice calculations in pure $SU(2)$ Yang-Mills \cite{GIACO}, though an eventual future result for $SU(3)$ is  desirable. 
 %%%%%%%%%%%%%%%%%%%%%%%%%%%%%%%%%%%%%%%
%\vspace*{-0.5cm}
%\vspace*{-0.3cm}
\section{Tests of the convergence of the OPE}
\label{sec:comments}
\vspace*{-0.25cm}
%%%%%%%%%%%%%%%%%%%%%%%%%%%%%%%%%%%%%%%
\nin
We show some behaviour of the OPE using the set of parameters obtained previously, namely the values of $\la\alpha_s G^2\ra$ and  $ \la g^3f_{abc} G^3\ra $  in Eqs. (\ref{eq:ag2}) and (\ref{eq:rhoc}) and the one of $\alpha_s$ in Eq. (\ref{eq:pt-param}). The PT series include the coefficient $c_3=-5.64$ of $\alpha_s^3$, while the $D=4$ condensate includes term to order $\alpha_s$. Representative expressions correspond to the moments where the optimal values of $\la\alpha_s G^2\ra$ and  $ \la g^3f_{abc} G^3\ra $ from Figs \ref{fig:ag2} and  \ref{fig:rho} are obtained. \\
\b Normalized to $(4m^2_c)^n\times 10^4$, the  ${\cal M}_{8,9}(4m_c^2)$ moments read: 
{\small
\bea
{\cal M}_9(4m_c^2)&=&  1.1314\ga 1.111-{0.407\over m_c^4}+{0.090\over m_c^6}+{0.085\over m_c^8}\dr, \nnb\\
{\cal M}_{10}(4m_c^2)&=& 0.4903\ga 1.013-{0.472\over m_c^4}+{0.152\over m_c^6}+{0.144\over m_c^8}\dr, 
\eea
}
while the  ${\cal M}_{15,16}(8m_c^2)$ moments normalized to $(4m^2_c)^n\times 10^9$ read:
{\small 
\bea
{\cal M}_{15}(8m_c^2)&=&2.3181\ga 1.005-{0.503\over m_c^4}+{0.112\over m_c^6}+{0.189\over m_c^8}\dr, \nnb\\
{\cal M}_{16}(8m_c^2)&=& 0.7077\ga 0.935-{0.549\over m_c^4}+{0.159\over m_c^6}+{0.263\over m_c^8}\dr, 
\label{eq:q2moments}
\eea
}
where one can see that the NP contributions become sizeable (the $\la\alpha_s G^2\ra$ contribution is 16-22\% of the LO contribution) but the OPE continues to converge (the $\la G^4\ra$ contribution is less than 4\%). Reciprocally, the relative large NP contributions have permitted the extraction of their size from the moments. \\
\b We also show the PT expressions of the moments normalized to $(4m^2_c)^n\times 10^4$ at fixed order:
{\small
\bea
{\cal M}^{PT}_9(4m_c^2) &=& 1.1314\ga1 + 0.601a_s + 2.7a_s^2 -  5.64a_s^3\dr~,\nnb\\
{\cal M}^{PT}_{10}(4m_c^2) &=&0.4903\ga 1 + 0.045a_s + 1.136a_s^2 - 5.64a_s^3\dr~,
\eea
 and (normalized to $(4m^2_c)^n\times 10^9$):
\bea
{\cal M}^{PT}_{15}(8m_c^2) &=&2.3181\ga 1 + 0.031a_s + 0.77a_s^2 - 5.64a_s^3\dr~,\nnb\\
{\cal M}^{PT}_{16}(8m_c^2) &=&0.7077\ga 1 - 0.364a_s - 0.33a_s^2 - 5.64a_s^3\dr~,
\eea
}
where $a_s\equiv \alpha_s/\pi$. 
One can note that radiative corrections to these higher moments are less than 11\% while it is about 30\% in the case of  ${\cal M}_0(0)$ in Eq. (\ref{eq:mom0}) which makes the latter sensitive to the way how the PT series is organized (fixed order, contour improved,...) as mentioned in~\cite{HOANG}.\\
\b The D=4 condensate contribution including the $\alpha_s$ corrections normalized to the LO PT moments and without the overall factor $\la a_s G^2\ra/(4\overline{m}_c^2)^2$ read:
{\small
\bea
{\cal M}^{D=4}_9(4m_c^2)&=&-329.4(1 - 0.862a_s)~,\nnb\\
{\cal M}^{D=4}_{10}(4m_c^2)&=&-433(1 - 1.673a_s)~,\nnb\\
{\cal M}^{D=4}_{15}(8m_c^2)&=&-413.8(1 - 0.986a_s)~,\nnb\\
{\cal M}^{D=4}_{16}(8m_c^2)&=&-491.3(1 - 1.527a_s)~.
\eea
}
Again here, the $\alpha_s$ corrections are relatively small which is not the case of  ${\cal M}_n(0)$ as one can see in Eq. (\ref{eq:g20}). 

 %%%%%%%%%%%%%%%%%%%%%%%%%%%%%%%%%%%%%%%
%\vspace*{-0.5cm}
%\vspace*{-0.3cm}
\section{Determinations of    $\overline{m}_{c,b}$ from low moments ${\cal M}_{n\leq 5}(0)$}
\label{sec:mass_0}
\vspace*{-0.25cm}
%%%%%%%%%%%%%%%%%%%%%%%%%%%%%%%%%%%%%%%
\nin
\b Low moments are widely used in the literature for extracting $\overline{m}_{c,b}$ where it has been
argued that its QCD expression is under a good control due to the negligible contributions of NPT terms.
Though this is absolutely true for  ${\cal M}_{1}(0)$, the neglect of the NPT terms becomes questionable for other moments because they increase in the OPE as shown explicitly in Eq. (\ref{eq:mom0}). The five lowest moments normalized to $(4m_c^2)^n$ read:
{\small
\bea
{\cal M}_1(0)&=& 0.8000\ga 1.300-{0.0222\over m_c^4}+{0.0005\over m_c^6}+{0.001\over m_c^8}\dr, \nnb\\
{\cal M}_2(0)&=& 0.3429\ga 1.350-{0.0862\over m_c^4}+{0.0076\over m_c^6}+{0.007\over m_c^8}\dr, \nnb\\
{\cal M}_3(0)&=&0.2023\ga 1.287-{0.1780\over m_c^4}+{0.0368\over m_c^6}+{0.027\over m_c^8}\dr, \nnb\\
{\cal M}_4(0)&=&0.1385\ga 1.158-{0.2815\over m_c^4}+{0.1172\over m_c^6}+{0.077\over m_c^8}\dr, \nnb\\
{\cal M}_5(0)&=&0.1023\ga 0.996-{0.3620\over m_c^4}+{0.2959\over m_c^6}+{0.184\over m_c^8}\dr, 
\label{eq:mom0}
\eea
}
which indicate that already for $n\geq 2$, one cannot neglect the non-perturbative contributions which are larger than 3.4\% (compared to  $\alpha_s^3 \geq 1.7\%$)
in the determination of $\overline{m}_c$. \\
\b Another inconvenience of
 ${\cal M}_{1}(0)$ is the large contribution ($\geq$ 40\% effect)  of
 the less accurate high-energy part of the spectral function which implies a model-dependent continuum contribution or a dependence on the way the non accurate data are handled as discussed explicitly in Section \ref{sec:cont}  and in \cite{HOANG}. \\
\b Low $Q^2=0$ moments are also affected by large radiative corrections which one can observe from their QCD expressions given in the literature  \cite{IOFFE,KUHN2,MAIER,MATEU,GREYNAT}.  To order $\alpha_s^3$, the PT series normalized to $(4\overline{m}_c^2)^n$ read in our normalization:
{\small 
 \bea
 {\cal M}^{PT}_1(0)&=&0.80(1+2.39a_s+2.38a_s^2-5.64a_s^3),\nnb\\
 {\cal M}^{PT}_2(0)&=&0.3429(1+2.43a_s+6.11a_s^2-7.64a_s^3),\nnb\\
 {\cal M}^{PT}_3(0)&=&0.2032(1+1.92a_s+6.12a_s^2-10.48a_s^3),\nnb\\
 {\cal M}^{PT}_4(0)&=&0.1385(1+1.10a_s+4.40a_s^2-18.13a_s^3),\nnb\\
 {\cal M}^{PT}_5(0)&=&0.1023(1+0.08a_s+2.16a_s^2-27.4a_s^3)~,
 \label{eq:pt0}
 \eea
 }
where one can notice that the coefficient of $a_s^3$ grows
with the order $n$ of the moments, but the coefficient of $\alpha_s$ decreases.  \\
\b The $D=4$ contribution including the $\alpha_s$ corrections normalized to the lowest order PT moments and without the overall factor $\la a_s G^2\ra/(4\overline{m}_c^2)^2$ read:
{\small
 \bea
 {\cal M}^{D=4}_1(0)&=&-15.04(1+2.48a_s),\nnb\\
  {\cal M}^{D=4}_2(0)&=&-58.49(1+1.05a_s),\nnb\\
   {\cal M}^{D=4}_3(0)&=&-143.6(1-0.48a_s),\nnb\\
     {\cal M}^{D=4}_4(0)&=&-283.4(1-2.11a_s),\nnb\\
 {\cal M}^{D=4}_5(0)&=&-491.3(1-3.80a_s)~,
 \label{eq:g20}
\eea
}
%%%%%%%%%%%%%%%%%%%%%%%%%%%%%%%
\begin{figure}[hbt]
\begin{center}
\includegraphics[width=7cm]{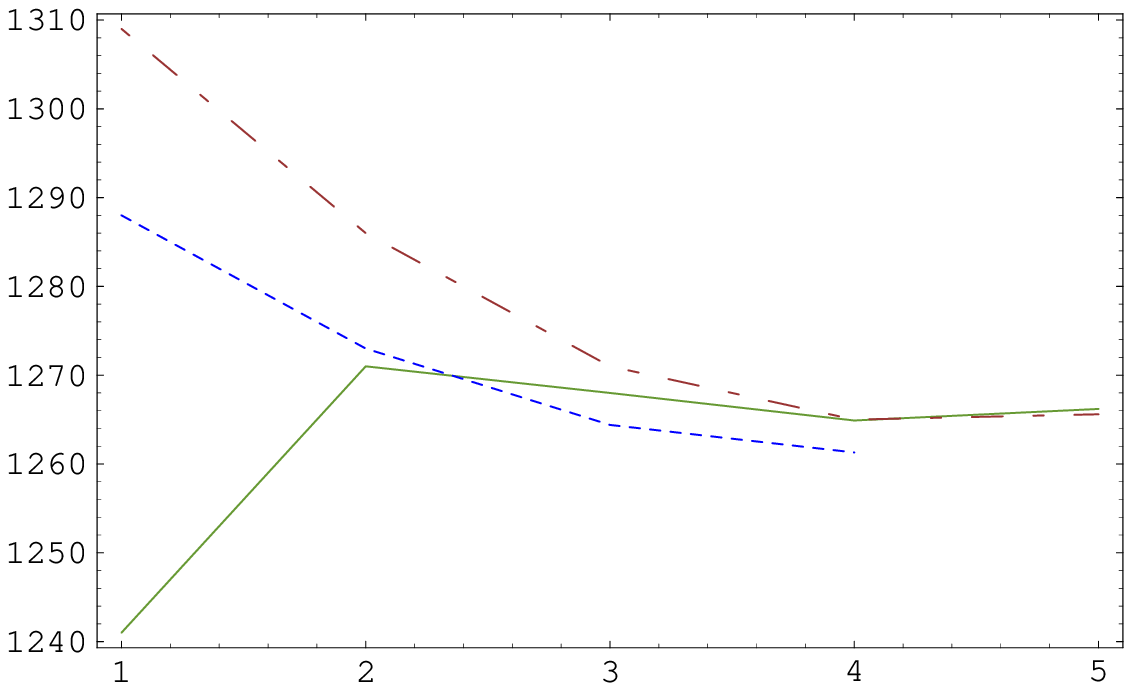}
\includegraphics[width=7cm]{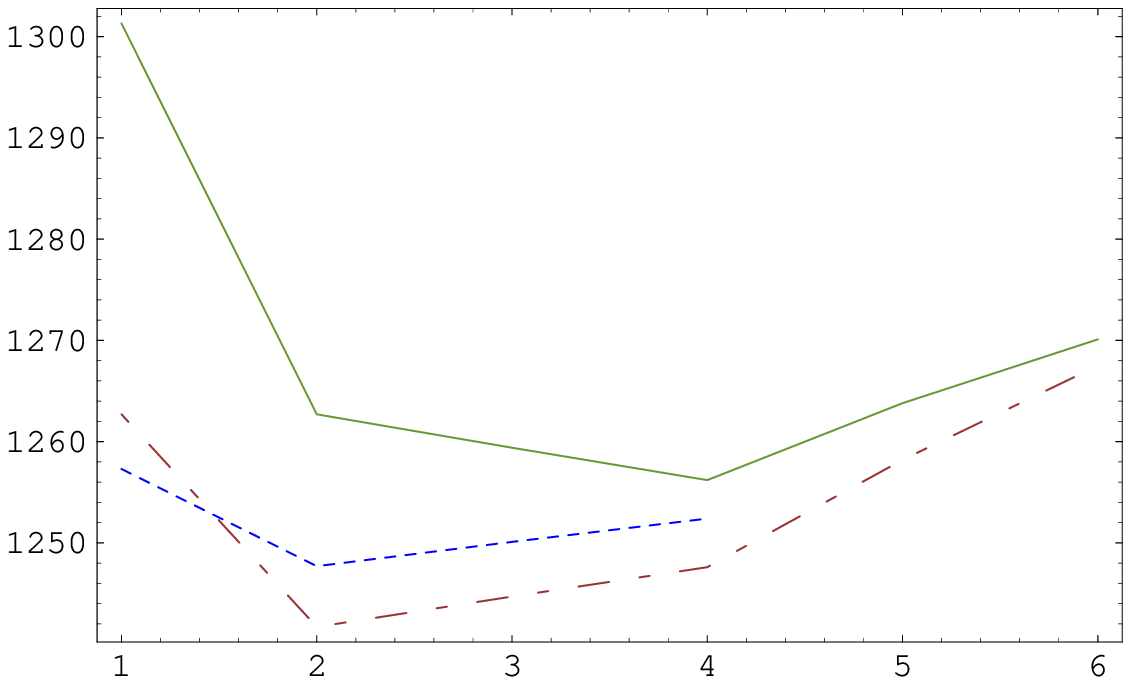}
%\vspace*{-0.7cm}
\caption{\footnotesize {\bf a)} Behaviour of $\overline{m}_c(\overline{m}_c)$  in units of MeV versus $n$ from ${\cal M}_n(0)$ moments   for the central value of $\la \alpha_s G^2\ra$ in Eq. (\ref{eq:ag2}) and of $\alpha_s$ given in Eq. (\ref{eq:alphas}). We use  fac=0.5 for the factorisation of $\la G^4\ra$.
%The region between the same curves correspond to the range of $\alpha_s$ values given in Eq. (\ref{eq:alphas}). 
The curves correspond to different models for the continuum defined in Section \ref{sec:cont} :    Model 1:  green (continued); Model 2:  red  (dot-dashed); Model 3 (data fit) :  blue (dot).
{\bf b)} Behaviour of $\overline{m}_c(\overline{m}_c)$  versus different ratio of moments  ${\cal M}_n(0)$ moments  in units of MeV. The inputs and legends are the same as in Fig \ref{fig:mass_0}a). In the $n$-axis: $1\equiv r_{1/2}, ~ 2\equiv r_{2/3},~ 3\equiv r_{2/4}, ~4\equiv r_{3/4}, ~5\equiv r_{3/5}$ and $6\equiv r_{4/5} $}
\label{fig:mass_0}
\end{center}
%\vspace*{-1.cm}
\end{figure} 
\nin
 %%%%%%%%%%%%%%%%%%%%%%%%%%%%%%%
where, one should note that one cannot go beyond $n=5$ because the $\alpha_s$ correction to the $D=4$ contribution is larger than 49\% indicating the divergence of the QCD series as also emphasized by \cite{IOFFE}. \\
\b Then, we limit ourselves to use the relatively low moments  ${\cal M}_{n\leq 5}(0)$
for extracting the running mass $\overline{m}_c(\overline{m}_c)$ within fixed order PT series and for a given set of NP parameters determined in the previous section. We show the results from the moments in Fig \ref{fig:mass_0}a) and the one from the ratios in Fig. \ref{fig:mass_0}b). As expected the result for $n\leq 2$ is sensitive to the Model for the continuum which contributes for 40\% to the moments. One can also note that using the moments from the data fit in \cite{HOANG} (Model 3), the result for $n=1$ is:
\beq
\overline{m}_c(\overline{m}_c)\vert^{1}_0=1289(8)~{\rm MeV}~,
\eeq
where the quoted error comes only from the change in $\alpha_s$ given in Eq. (\ref{eq:alphas}) (some other sources of errors will be discussed later on). Though this result agrees with different determinations from ${\cal M}_1(0)$ \cite{KUHN2,MAIER,HOANG,KUHN1}, one can note that its central value decreases when one increases the number of derivative $n$ of the moments. The result only stabilizes versus the variation of $n$ for $n\geq 3-5$, where an optimal result can be taken. For definiteness, we take $n\simeq 4$ , where all {\it Continuum Models} give consistent predictions. Then, we deduce, from Fig \ref{fig:mass_0}a),  in units of MeV:
\bea
\overline{m}_c(\overline{m}_c)\vert^4_0=1263.7&&(1.3)_{cont}(3.5)_{\alpha_s}(4.9)_{\alpha^4_s}(3.9)_{\mu}\nnb\\
&&(4.4)_{G^2}(4.7)_{G^3}(3.1)_{G^4}(1.7)_{exp}~,
\label{eq:mass_0a}
\eea
where the central value is the average from different continuum models.  It leads to the result in Table \ref{tab:resc}.\\
%\beq
%\overline{m}_c(\overline{m}_c)\vert_{0}=1263.7(10.2)~{\rm MeV}.
%\label{eq:mass_0}
%\eeq
\b The 1st error in Eq. (\ref{eq:mass_0a}) is due to the different models for the continuum, the 2nd one  to the value of $\alpha_s$ given in Eq. (\ref{eq:alphas}). The 3rd error is an estimate of higher order terms of PT assumed to be  equal to the contribution of the $\alpha_s^3$ one, while the 4th error is  an estimate of the effect of the subtraction point $\nu$ by varying it from $m_c$ to $M_\tau$ and using the substitution (see e.g. \cite{SNB3,SNB1}):
\bea
\alpha_s(m_c)&\to& \alpha_s(\nu)\times \ga 1-{\beta_1}~{\alpha_s(\nu)\over\pi}\log{\nu\over  m_c}\dr~,
%{\cal M}^{(2)}_{PT}&\to&{\cal M}^{(2)}_{PT}+{\cal M}^{(1)}_{PT}~\beta_1\log{\nu\over  m_c}~,
\label{eq:sub}
\eea
where $\beta_1=-(1/2)(11-2n_f/3)$ for $n_f$-flavours. The 5th and 6th errors are due respectively to the ones of the gluon condensates $\la\alpha_s G^2\ra$  and $\la g^3f_{abc}G^3\ra$ estimated previously. The 7th error is due to the $\la G^4\ra$ condensates allowing it to move from fac=0.5 (favoured value from our preceeding fit) to fac=1 as defined in Eq. (\ref{eq:fac}). The last error is due to the experimental $J/\psi$ widths given in Table \ref{tab:psi}. \\
\b We consider as a final value the one obtained from ${\cal M}_{4}(0)$ where both PT corrections are still small for the unit and dimension 4 operators. Indeed, for the unit operator, the dominant correction is due to $\alpha_s$, which  is about 14\% for ${\cal M}_{4}(0)$ or for $m_c^2$\,\footnote{ The $\alpha_s^2$ (resp $\alpha_s^3$) are relatively small i.e 7.4\% (resp. 3.9\%).} and which is about half of the one of ${\cal M}_{1}(0)$. Then, we may expect that the error induced by the organization of the PT series (fixed order, contour improved,...) discussed in \cite{HOANG} is smaller for ${\cal M}_{4}(0)$ though the PT series converges faster for ${\cal M}_{1}(0)$ as on can notice in Eq. (\ref{eq:pt0}). 
\\
\b The result from the ratios of moments in Fig \ref{fig:mass_0}b)  is not very conclusive as the model-dependence of the result starts to disappear  from the ratio of moments $r_{3/5}$, but for these ratios the  result increases with $n$. Then, we shall not retain the results from the ratios of moments for the charmonium channel.
 %%%%%%%%%%%%%%%%%%%%%%%%%%%%%%%%%%%%%%%
%\vspace*{-0.5cm}
%\vspace*{-0.3cm}
\section{ $\overline{m}_{c}(\overline{m}_{c})$ from higher ${\cal M}_{n}(Q^2)$ moments }
\label{sec:mass_4m2}
\vspace*{-0.25cm}
%%%%%%%%%%%%%%%%%%%%%%%%%%%%%%%%%%%%%%%
\nin
%%%%%%%%%%%%%%%%%%%%%%%%%%%%%%%
\begin{figure}[hbt]
\begin{center}
\includegraphics[width=7cm]{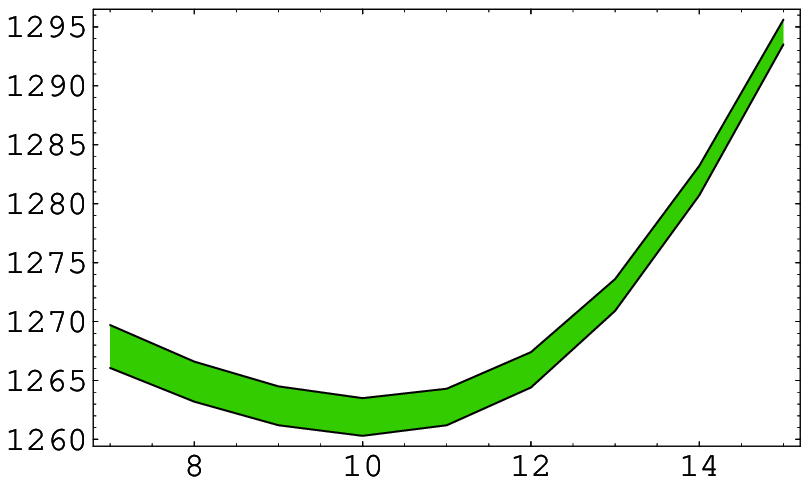}
\includegraphics[width=7cm]{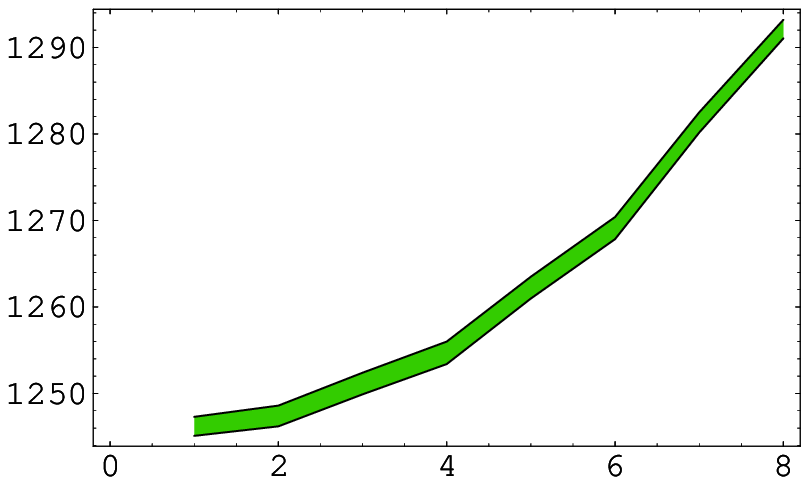}
%\vspace*{-0.7cm}
\caption{\footnotesize  {\bf a)} Behaviour of $\overline{m}_c(\overline{m}_c)$ in units of MeV  versus $n$ from ${\cal M}_n(4\overline{m_c}^2)$ moments  and  for the central value of $\la \alpha_s G^2\ra$ in Eq. (\ref{eq:ag2}). We use  fac=0.5 for the factorisation of $\la G^4\ra$ and Model 2 for the continuum. 
The colored region corresponds to the range of $c_3$ values given in Eq. (\ref{eq:as3}). {\bf b)} Behaviour of $\overline{m}_c(\overline{m}_c)$ in units of MeV versus different ratios of moments  $r_{n/n+1}(4\overline{m_c}^2)$  . The inputs and legends are the same as in Fig \ref{fig:mass_4m2}a). In the $n$-axis: $1\equiv r_{7/9}, ~ 2\equiv r_{8/9},~ 3\equiv r_{8/10}, ~4\equiv r_{9/10}$,\dots}
\label{fig:mass_4m2}
\end{center}
%\vspace*{-1.cm}
\end{figure} 
\nin
 %%%%%%%%%%%%%%%%%%%%%%%%%%%%%%%
\begin{figure}[hbt]
\begin{center}
\includegraphics[width=7cm]{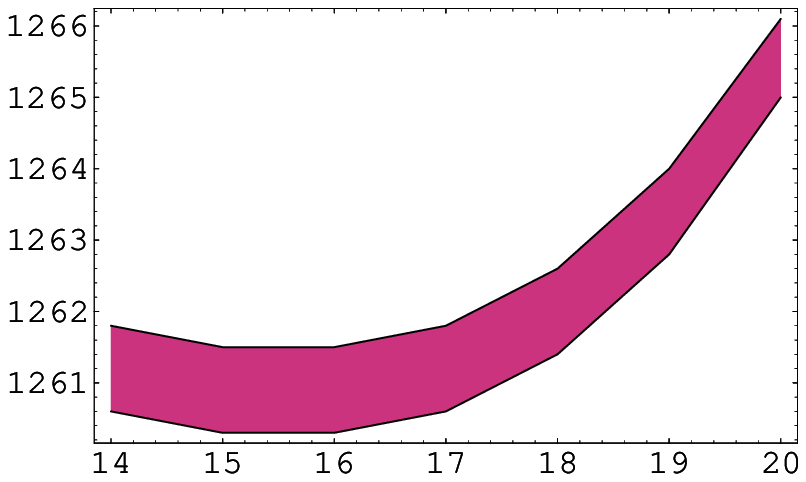}
\includegraphics[width=7cm]{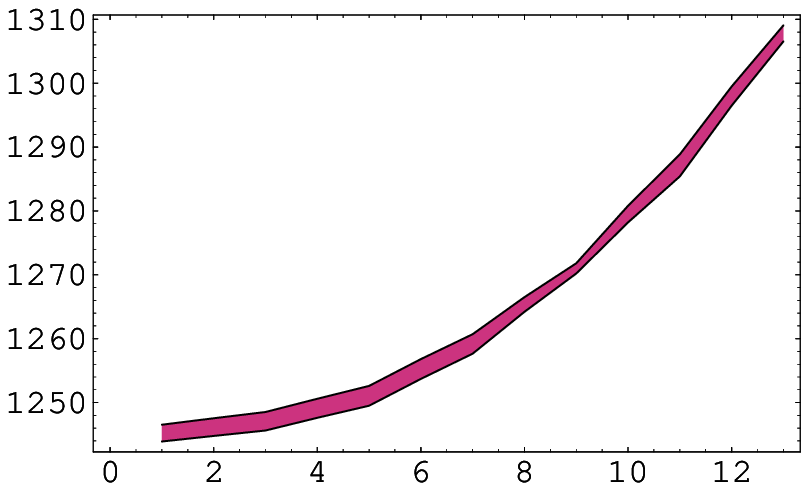}
%\vspace*{-0.7cm}
\caption{\footnotesize  {\bf a)} Behaviour of $\overline{m}_c(\overline{m}_c)$ in units of MeV  versus $n$ from ${\cal M}_n(8\overline{m_c}^2)$ moments  for the central value of $\la \alpha_s G^2\ra$ in Eq. (\ref{eq:ag2}) and of $\alpha_s$ given in Eq. (\ref{eq:alphas}). We use  fac=0.5 for the factorisation of $\la G^4\ra$ and Model 2 for the continuum. 
The colored region corresponds to the range of $c_3$ values given in Eq. (\ref{eq:as3}).  {\bf b)} Behaviour of $\overline{m}_c(\overline{m}_c)$  versus different ratios of moments  ${\cal M}_n(8\overline{m_c}^2)$  in units of MeV. The inputs and legends are the same as in Fig \ref{fig:mass_8m2}a). In the $n$ axis: $1\equiv r_{13/14}, ~ 2\equiv r_{13/15},~ 3\equiv r_{14/15}, ~4\equiv r_{14/16}, ~5\equiv r_{15/16}$,\dots}
\label{fig:mass_8m2}
\end{center}
%\vspace*{-1.cm}
\end{figure} 
\nin
 %%%%%%%%%%%%%%%%%%%%%%%%%%%%%%%
\b  In the following, we shall extract $\overline{m}_{c}(\overline{m}_{c})$ from higher ${\cal M}_{n}(4m_c^2)$ and ${\cal M}_{n}(8m_c^2)$ moments. We show the results of the analysis respectively from the moments in Figs \ref{fig:mass_4m2}a) and \ref{fig:mass_8m2}a) and from the ratios of moments in Figs \ref{fig:mass_4m2}b) and \ref{fig:mass_8m2}b). One can notice that in both cases the results from the moments present minimas versus $n$. \\
\b The  minimum is obtained from ${\cal M}_{10}(4m_c^2)$ and from  ${\cal M}_{15}(8m_c^2)$, which give in units of MeV:
\bea
\overline{m}_{c}(\overline{m}_{c})\vert^{10}_{4m_c^2}=1261.9&&(0.7)_{\alpha_s}(1.6)_{\alpha_s^3}(1.6)_{\alpha_s^{n\geq 4}}(0.4)_\mu\nnb\\
&&(1.1)_{G^2}(1.0)_{G^3}(1.7)_{G^4}(3.0)_{exp}~,\nnb\\
\overline{m}_{c}(\overline{m}_{c})\vert^{15}_{8m_c^2}=1260.9&&(0.5)_{\alpha_s}(1.6)_{\alpha_s^3}(1.6)_{\alpha_s^{n\geq 4}}(0.4)_\mu\nnb\\
&&(1.0)_{G^2}(0.7)_{G^3}(1.3)_{G^4}(2.6)_{exp}~,\nnb\\
\label{eq:mass_4m2a}
\eea
which lead to the result in Table \ref{tab:resc}.
The different sources of errors are the same as the ones discussed in Eq. (\ref{eq:mass_0a}). The one from ${\alpha_s^3}$ here is due to the distance of the average of the $\alpha_s^3$ contribution to the $\pm$ assumed value of the coefficient  in Eq. (\ref{eq:as3}). We have estimated the error due to the unknown  $\alpha_s^n~(n\geq 4)$ to be equal to that of $\alpha_s^3$. Eq. (\ref{eq:mass_4m2a}) leads to the result in Table~\ref{tab:resc}.\\
\b One can also see in Figs \ref{fig:mass_4m2}b) and \ref{fig:mass_8m2}b) that the results from the ratios of moments increase with $n$. Though, the outputs obtained from the ratios of optimal moments are consistent with the ones from these moments
and with the ones obtained in \cite{SNcb} where a judicious choice (small PT corrections) of these ratios have been used, we shall not consider these numbers in the final results because of the absence of stabilities or minimas versus the variation of n. 
%%%%%%%%%%%%%%%%%%%%%%%%%%%%%%%
 %\vspace*{-0.5cm}
{\scriptsize
\begin{table}[hbt]
\setlength{\tabcolsep}{3pc}
 \caption{\scriptsize    Value of $\overline{m}_c(\overline m_c)$ from charmonium moments known to $\alpha_s^3$ for $Q^2=0$ and with an estimate of the $\alpha_s^3$ contribution for $Q^2\not= 0$.}
{\small
\begin{tabular}{ll}
&\\
\hline
%\hline
%\\
Moments&$\overline{m}_c(\overline m_c)$ [MeV] \\
%\\
\hline
%\hline
\\
$Q^2$=0:\\
${\cal M}^4$&$1263.7(10.3)$\\
\\
$Q^2$=$4m_c^2:$ \\
${\cal M}^{10}$&$1261.9(4.5)$\\
\\
$Q^2$=$8m_c^2:$\\
${\cal M}^{15}$&$1260.9(4.0)$\\
\\
\hline
%Average &1262.1(6.3)\\
%\hline
\end{tabular}
}
\label{tab:resc}
\end{table}
}
%\vspace*{-0.2cm}
\nin
%%%%%%%%%%%%%%%%%%%%%%%%%%%%%%%%%
 %%%%%%%%%%%%%%%%%%%%%%%%%%%%%%%%%%%%%%%
%\vspace*{-0.5cm}
%\vspace*{-0.3cm}
\section{ Final value of $\overline{m}_{c}(\overline{m}_{c})$ and Coulombic corrections }
\label{sec:final_mc}
\vspace*{-0.25cm}
%%%%%%%%%%%%%%%%%%%%%%%%%%%%%%%%%%%%%%%
\nin
\b Like in \cite{SNcb}, we  approximately estimate the Coulombic correction by working with the resummed expression of the spectral function \cite{EICHTEN}\,\footnote{However, we (a priori) expect that the non-relativisitc corrections will be relatively small here as we are working in the relativistic domain because $m_c$ is relatively light, while the final result corresponds to a relatively large $Q^2=8m_c^2$ value.} :
 \beq
 {\cal R}_c\vert_{Coul}\simeq {3\over 2}v{x\over 1-e^{-x}}~,
  \label{eq:coulomb2}
 \eeq
 in an expansion in terms of $x\equiv  C_F\pi\alpha_s/v$ where $C_F=4/3$ and $v$ is the quark velocity defined in Eq. (\ref{eq:velocity}). This contribution, which is of long-distance origin and proportional to the imaginary part (the wave function) of the two-point function, is induced by rescattering (Sommerfeld factor) of heavy quark pairs through the Coulomb potential above the $\bar cc$ threshold\,\footnote{The Coulomb corrections arising from the bound states
below the threshold can be safely neglected as the dispersion relation is applied
above threshold $(t\geq 4m_c^2)$ where the QCD expression of the spectral function from field theory
(OPE) is used while its phenomenological expression is measured from the $e^+e^-\to $ hadrons data.}. \\
\b We add to this expression some PT QCD corrections. The 1st correction is the familiar $(1-4C_Fa_s)$ factor due to the quarkonium annihilation through a single (transverse) virtual gluon. The 2nd type of corrections to order $v$ and $\log v$ have been obtained in \cite{CHET1,CHET2} up to order $\alpha_s^2$ where the result is strictly applicable near threshold $C_F\pi\alpha_s\leq v\ll 1$\,\footnote{However, according to Refs. \cite{CHET1,CHET2}, these short-distance effects being specific for the single annihilation process involving $\bar QQ$ pairs are universal for $|v|\ll 1$ regardless  whether $|v|$ is smaller or larger than  $C_F\pi\alpha_s$.}. \\
\b  We compare the value of the moments using the previous expressions for the spectral function with the one obtained from PT theory including radiative corrections up to order $\alpha_s^2$. In the case $Q^2= 8m_c^2$ and $n=15$, where the most precise result is obtained, the corrections induced by the previous Coulombic contributions to the value of $m_c$ is about -1.3\%\,\footnote{Some further arguments justifying a much smaller value of these contributions can be found in \cite{IOFFE}. A much smaller effect of about  1 MeV is also obtained for the ratio of moments like has been found in \cite{SNcb}. }~\footnote{The effect on $\overline{m}_{c}(\overline{m}_{c})$ from ${\cal M}^{10}(4m_c^2)$ and ${\cal M}^4(0)$ are respectively 2 \% and 5\%.} and gives:
\beq
 \delta_{m_c}\vert_{Coul}=\pm 16~{\rm MeV}~.
 \label{eq:coulomb}
 \eeq 
 We consider this effect as another source of error rather than a definite shift on $\overline{m}_{c}(\overline{m}_{c})$ due to the fact 
 that the r\^ole of the Coulombic  effect in the sum rule analysis remains unclear \cite{IOFFE,NOVIKOV2} as the quark is still relativistic with a relatively large velocity:
 \beq
 v\approx \sqrt{\ga 1+Q^2/4m_Q^2\dr/n}\simeq 0.45~,
\eeq
for large $n=15$ and $Q^2=8m_Q^2$. Indeed, this value of $v$ would correspond to a momentum transfer between quark and anti-quark of about 1 GeV, where the effective potential differs from the Coulombic one \cite{EICHTEN} and where the sum rules are usually successfully applied. 
  \\
   \b One can see from Table \ref{tab:resc} that the estimate from different forms of the moments are consistent
each other. We shall consider as a final estimate the most precise one from ${\cal M}^{15}(8m_c^2)$,
where the Coulombic correction obtained previously is also small. Adding this correction,
we obtain:
\beq
\overline{m}_{c}(\overline{m}_{c})=1261(16)~{\rm MeV}~,
\label{eq:mass_mc}
\eeq
in excellent agreement with the one \cite{SNcb}:
\beq
\overline{m}_{c}(\overline{m}_{c})=1261(18)~{\rm MeV}~,
\label{eq:mass_mca}
\eeq
obtained from a judicious choice of ratios of high moments having small PT and NP corrections. The previous results also improve earlier results obtained
by the author to lower orders in this channel \cite{SNmc}.

%%%%%%%%%%%%%%%%%%%%%%%%%%%%%%%%
\section{ Determination of  \boldmath$\overline{m}_b(\overline m_b)$}
\vspace*{-0.25cm}
%%%%%%%%%%%%%%%%%%%%%%%%%%%%%%%%%%%%%%%
 \nin
 We extend the previous analysis  to the bottomium systems. In the following, we shall use 
 the value:
 \beq
 \alpha_s(m_b)\vert_{n_f=5}=0.219(4)~,
 \eeq 
 deduced from $\alpha_s(m_\tau)$ in Eq. (\ref{eq:alphas}).  
  %%%%%%%%%%%%%%%%%%%%%%%%%%%%%%%
% \vspace*{-1.cm}
{\scriptsize
\begin{table}[hbt]
\setlength{\tabcolsep}{1.6pc}
 \caption{\scriptsize    Masses and electronic widths of the  $\Upsilon$ family from PDG10\cite{PDG}. }
{\small
\begin{tabular}{lll}
&\\
\hline
%\hline
%\\
Name&Mass [MeV]&$\Gamma_{\Upsilon\to e^+e^-}$ [keV] \\
%\\
\hline
%\hline
\\
$\Upsilon(1S)$&9460.30(26)&1.340(18)\\
$\Upsilon(2S)$&10023.26(31)&0.612(11)\\
$\Upsilon(3S)$&10355.2(5)&0.443(8)\\
$\Upsilon(4S)$&10579.4(1.2)&0.272(29)\\
$\Upsilon(10860)$&10865(8)&0.31(7)\\
$\Upsilon(11020)$&11019(8)&0.13(3)\\
\\
\hline
\end{tabular}
}
\label{tab:upsilon}
\end{table}
}
%\vspace*{-2.5cm}
\nin
%%%%%%%%%%%%%%%%%%%%%%%%%%%%%%%%%
  %%%%%%%%%%%%%%%%%%%%%%%%%%%%%%%
%%%%%%%%%%%%%%%%%%%%%%%%%%%%%%%
\begin{figure}[hbt]
\begin{center}
\includegraphics[width=7cm]{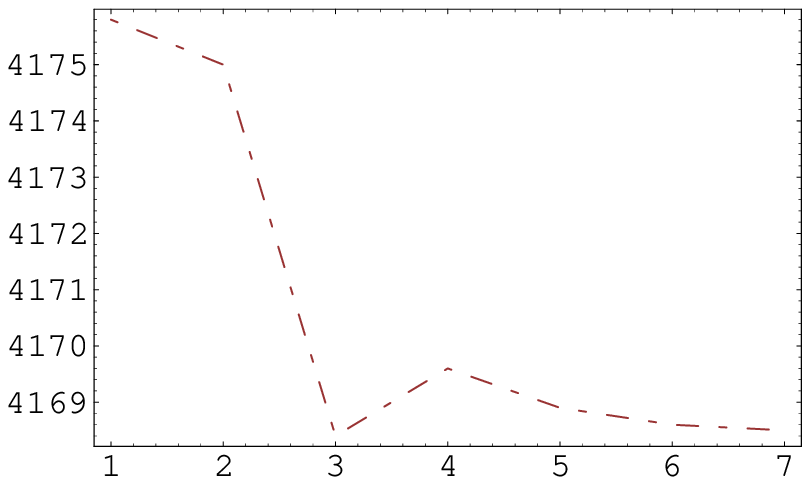}
\includegraphics[width=7cm]{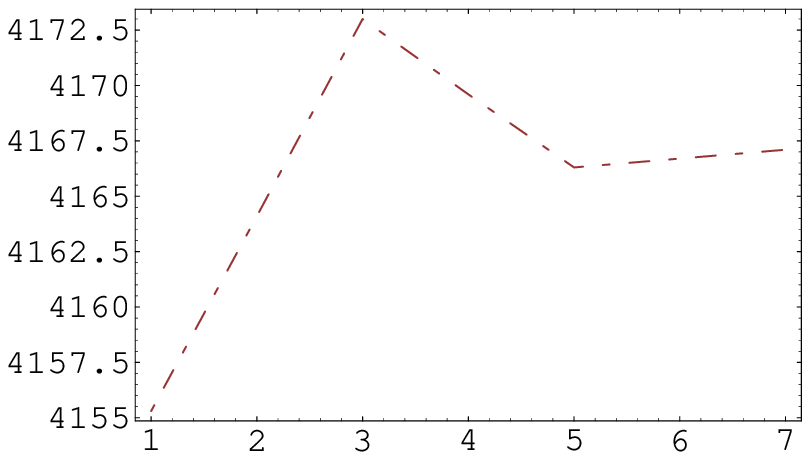}
%\vspace*{-0.7cm}
\caption{\footnotesize  {\bf a)} Behaviour of $\overline{m}_b(\overline{m}_b)$ in units of MeV  versus $n$ from ${\cal M}_n(0)$ moments  for the central value of $\la \alpha_s G^2\ra$ in Eq. (\ref{eq:ag2}) and of $\alpha_s$ given in Eq. (\ref{eq:alphas}).
We use  fac=0.5 for the factorisation of $\la G^4\ra$ and Model 2 for the high-energy part of the spectral function.  {\bf b)} Behaviour of $\overline{m}_b(\overline{m}_b)$  versus different ratios of moments  ${\cal M}_n(0)$  in units of MeV. The inputs and legends are the same as in Fig \ref{fig:mb_0}a). In the $n$-axis: $1\equiv r_{2/3}, ~ 2\equiv r_{2/4},~ 3\equiv r_{3/4}, ~4\equiv r_{3/5}, ~5\equiv r_{4/5}$,\dots}
\label{fig:mb_0}
\end{center}
%\vspace*{-1.cm}
\end{figure} 
\nin
 %%%%%%%%%%%%%%%%%%%%%%%%%%%%%%%
%%%%%%%%%%%%%%%%%%%%%%%%%%%%%%%
\begin{figure}[hbt]
\begin{center}
\includegraphics[width=7cm]{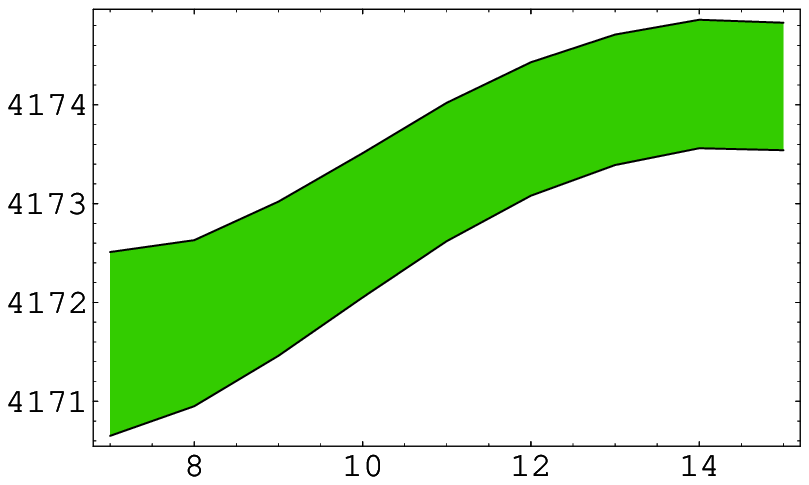}
\includegraphics[width=7cm]{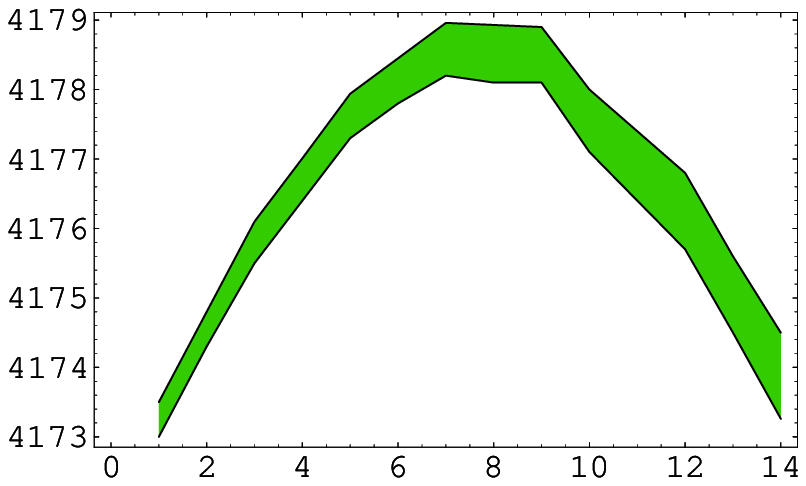}
%\vspace*{-0.7cm}
\caption{\footnotesize  {\bf a)} Behaviour of $\overline{m}_b(\overline{m}_b)$ in units of MeV  versus $n$ from ${\cal M}_n(4\overline{m_b}^2)$ moments for the central value of $\la \alpha_s G^2\ra$ in Eq. (\ref{eq:ag2}) and of $\alpha_s$ given in Eq. (\ref{eq:alphas}).
 We use  fac=0.5 for the factorisation of $\la G^4\ra$ and Model 2 for the high-energy part of the spectral function. The colored region corresponds to the range of $c_3$ values given in Eq. (\ref{eq:as3}).
{\bf b)} Behaviour of $\overline{m}_b(\overline{m}_b)$  versus different ratios of moments  ${\cal M}_n(4\overline{m_b}^2)$  in units of MeV. The inputs and legends are the same as in Fig \ref{fig:mb_4m2}a). In the $n$ axis: $1\equiv r_{7/8},~ 2\equiv r_{7/9}, ~3\equiv r_{8/9}, ~4\equiv r_{8/10}$, \dots}
\label{fig:mb_4m2}
\end{center}
%\vspace*{-1.cm}
\end{figure} 
\nin
%%%%%%%%%%%%%%%%%%%%%%%%%%%%%%%
\begin{figure}[hbt]
\begin{center}
\includegraphics[width=7cm]{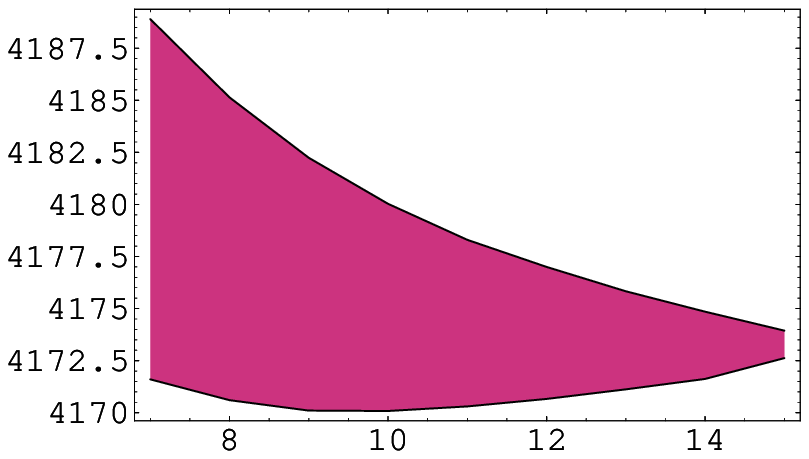}
\includegraphics[width=7cm]{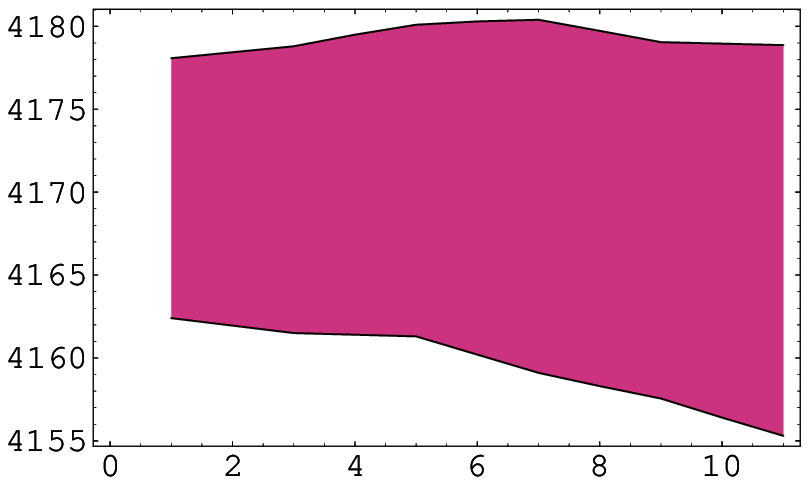}
%\vspace*{-0.7cm}
\caption{\footnotesize  {\bf a)} Behaviour of $\overline{m}_b(\overline{m}_b)$ in units of MeV  versus $n$ from ${\cal M}_n(8\overline{m_b}^2)$ moments for the central value of $\la \alpha_s G^2\ra$ in Eq. (\ref{eq:ag2}) and and of $\alpha_s$ given in Eq. (\ref{eq:alphas}). We use  fac=0.5 for the factorisation of $\la G^4\ra$.
Model 2 for the high-energy part of the spectral function is used. The colored region corresponds to the range of $c_3$ values given in Eq. (\ref{eq:as3}).
{\bf b)} Behaviour of $\overline{m}_b(\overline{m}_b)$  versus different ratios of moments  ${\cal M}_n(8\overline{m_b}^2)$  in units of MeV. The inputs and legends are the same as in Fig \ref{fig:mb_8m2}a). In the $n$ axis: $1\equiv r_{13/14},~ 2\equiv r_{13/15}, ~4\equiv r_{14/15}$, \dots}
\label{fig:mb_8m2}
\end{center}
%\vspace*{-1.cm}
\end{figure} 
\nin
%%%%%%%%%%%%%%%%%%%%%%%%%%%%%%%
 %\vspace*{-0.5cm}
{\scriptsize
\begin{table}[hbt]
\setlength{\tabcolsep}{3pc}
 \caption{\scriptsize    Value of $\overline{m}_b(\overline m_b)$ from bottomium moments known to $\alpha_s^3$ for $Q^2=0$ and with an estimate of the $\alpha_s^3$ contribution for $Q^2\not= 0$.}
{\small
\begin{tabular}{ll}
&\\
\hline
%\hline
%\\
Moments&$\overline{m}_b(\overline m_b)$ [MeV] \\
%\\
\hline
%\hline
\\
$Q^2$=0:\\
${\cal M}^4$&$4169.6(11.3)$\\
$r_{4/5}$&$4166.4(18.6)$\\
\\
$Q^2$=$4m_b^2:$ \\
${\cal M}^{14}$&$4174.2(5.8)$\\
$~r_{10/11}$&$4178.6(12.1)$\\
\\
$Q^2$=$8m_b^2:$\\
${\cal M}^{10}$&$4175.1(5.5)$\\
$r_{16/17}$&$4170.9(12.7)$\\
\\
\hline
%Average &4172.5(11.0)\\
%\hline
\end{tabular}
}
\label{tab:resb}
\end{table}
}
%\vspace*{-0.2cm}
\nin
%%%%%%%%%%%%%%%%%%%%%%%%%%%%%%%%%
We shall use as experimental inputs the $\Upsilon$-family parameters in Table \ref{tab:upsilon} using NWA and parametrize the 
spectral function above $\sqrt{t}=(11.098\pm 0.079)$ GeV by its pQCD expression (Model 2), where the error in the continuum  threshold is given by the total width of the $\Upsilon(11020)$.
We shall work with higher moments  in order to minimize the contributions of the  QCD continuum.  
%the dominant contributions will come from the two lowest ground states while 
%finite width corrections will not be observable.  
We use 
moments known to order $\alpha_s^3$ for $Q^2=0$, while for $Q^2\not=0$,  we have added the estimate of the $\alpha_s^3$ contribution given in Eq. (\ref{eq:as3}). 
 %%%%%%%%%%%%%%%%%%%%%%%%%%%%%%%%%%%
 \subsection*{\b Results from ${\cal M}_n(0)$}
\nin
%%%%%%%%%%%%%%%%%%%%%%%%%%%%%%%%%%%
 We show the results from ${\cal M}_n(0)$ in Fig \ref{fig:mb_0}, where one can notice that the result is (almost) stable versus the  variation of $n$  for $n\simeq 3\sim 7$ while for the ratios of moments, the stability is reached from $r_{4/5}$. At these values, the contribution of the QCD continuum is less than 29\% of the total which is much less than the one for $n=1$ where it is about 66\%. This feature raises serious doubts on the accurate value of $m_b$ from this low moment   ${\cal M}_1(0)$ given in the literature \cite{KUHN2,MAIER,KUHN1} due to the inaccuracy of the data in this high-energy region. From the moments, we obtain in units of MeV:
 \bea
\overline{m}_{b}(\overline{m}_{b})\vert^{4}_{0}=4169.6&&(1.9)_{\alpha_s}(4.1)_{\alpha_s^{n\geq 4}}(2.7)_\mu\nnb\\
&&(1.1)_{G^2}(1.2)_{G^3}(1.2)_{G^4}(10.6)_{exp}~,\nnb\\
\overline{m}_{b}(\overline{m}_{b})\vert^{4/5}_{0}=4166.3&&(4)_{\alpha_s}(7.1)_{\alpha_s^{n\geq 4}}(16.6)_\mu\nnb\\
&&(0.6)_{G^2}(0.4)_{G^3}(0.4)_{G^4}(1.9)_{exp}~,\nnb\\
\label{eq:mb_0}
\eea
giving the results in Table \ref{tab:resb}. One can notice that, at the optimal choice $r_{4/5}(0)$, PT corrections are large which induce larger PT errors than in the case of ${\cal M}_4(0)$. The different sources of errors are similar to the case of charmonium.
 %%%%%%%%%%%%%%%%%%%%%%%%%%%%%%%%%%%
 \subsection*{\b Results from ${\cal M}_n(4m_b^2)$}
\nin
%%%%%%%%%%%%%%%%%%%%%%%%%%%%%%%%%%%
The results from ${\cal M}_n(4m_b^2)$ are shown in Fig \ref{fig:mb_4m2}, where a stability versus the variation of $n$ is obtained for $n=14$, while for the ratios of moments, it is reached for $r_{10/11}(4m_b^2)$ and $r_{10/12}(4m_b^2)$.  In both cases, the errors due to the NP contributions and induced by the $\pm$ sign for the estimate of the $\alpha_s^3$ coefficient  and of the higher order $\alpha_s^{n\geq 4}$ are tiny ($\leq 0.4$ MeV) and can be neglected. We obtain in units of MeV:
\bea
\overline{m}_{b}(\overline{m}_{b})\vert^{14}_{4m_b^2}&=&4174.2(0.6)_{\alpha_s}(2.6)_\mu(5.1)_{exp} \nnb\\
\overline{m}_{b}(\overline{m}_{b})\vert^{10/11}_{4m_b^2}&=&4178.6(4.2)_{\alpha_s}(10.8)_\mu(3.6)_{exp},
\label{eq:mb_4m2}
\eea
from which, we deduce the result in Table \ref{tab:resb}.  
 %%%%%%%%%%%%%%%%%%%%%%%%%%%%%%%%%%%
 \subsection*{\b Results from ${\cal M}_n(8m_b^2)$}
\nin
%%%%%%%%%%%%%%%%%%%%%%%%%%%%%%%%%%%
The results from ${\cal M}_n(8m_b^2)$ and from the ratios of moments are shown in Fig \ref{fig:mb_8m2}, where stabilities versus the $n$-variations are respectively obtained for $n=9\sim 11$ and for $r_{15/17}, ~r_{16/17}$. Non perturbative corrections and the one due to the $\pm$ sign of the $\alpha_s^3$ coefficient are also negligible ($\leq 0.3$ MeV). We obtain in units of MeV:
\bea
\overline{m}_{b}(\overline{m}_{b})\vert^{10}_{8m_b^2}=4175.1&&(0.5)_{\alpha_s}(5.1)_{\alpha_s^3}(0.5)_{\alpha_s^{n\geq 4}}\nnb\\
&&(1.9)_\mu(10)_{exp}~,\nnb\\
\overline{m}_{b}(\overline{m}_{b})\vert^{16/17}_{8m_b^2}=4170.9&&(1.6)_{\alpha_s}(9.5)_{\alpha_s^3}(4.1)_{\alpha_s^{n\geq 4}}\nnb\\
&&(7.2)_\mu(3.6)_{exp}~.
\label{eq:mb_8m2}
\eea
Then, we deduce the result in Table \ref{tab:resb}.  
%%%%%%%%%%%%%%%%%%%%%%%%%%%%%%%%%%%%%%%
%\vspace*{-0.5cm}
%\vspace*{-0.3cm}
\section{ Final value of $\overline{m}_{b}(\overline{m}_{b})$ and Coulombic corrections }
\label{sec:final_mb}
\vspace*{-0.25cm}
%%%%%%%%%%%%%%%%%%%%%%%%%%%%%%%%%%%%%%%
\nin
\b
Here, we analyze the Coulombic corrections like in the case of charm. The ones for the moments are
relatively large which are respectively 1.7\%, 1.1\% and 4\% for ${\cal M}^{10}(8m_b^2)$, ${\cal M}^{14}(4m_b^2)$ and ${\cal M}^4(0)$. The ones for the ratios of moments $r_{16/17}(8m_b^2)$, $~r_{10/11}(4m_b^2)$ and $r_{4/5}(0)$ are respectively 1.2, 2. 1 and 3.6 per mil, which are about one order of magnitude smaller.  Among these different
determinations the one from $r_{16/17}(8m_b^2)$ is the most precise. We consider it as our best final result:
\beq
\overline{m}_{b}(\overline{m}_{b})=4171(14)~{\rm MeV}.
\label{eq:mass_mb}
\eeq
It is informative to compare the previous result with the one in \cite{SNcb} (see Table  \ref{tab:resb0} from \cite{SNcb}) using some judicious choices of the ratios of moments having the smallest PT corrections and where the $\la G^4\ra$ contribution has not been included. 
%After correcting a bug in the program used in \cite{SNcb} for the continuum contribution, one obtains the average from Table \ref{tab:resb0}. 
Adding the errors $\pm 6$ MeV due to the Coulombic, $\pm 6$ MeV due to the subtraction point and $\pm 4$ MeV due to the $\alpha_s^3$ contributions, the average from Table \ref{tab:resb0} becomes:
\bea
\overline{m}_{b}(\overline{m}_{b})&=&4173(10)~{\rm MeV}~,
\eea
which is in excellent agreement with the one obtained in Eq. (\ref{eq:mass_mb}). 
%%%%%%%%%%%%%%%%%%%%%%%%%%%%%%%
 %\vspace*{-0.5cm}
{\scriptsize
\begin{table}[hbt]
\setlength{\tabcolsep}{3pc}
 \caption{\scriptsize    Corrected values of $\overline{m}_b(m_b)$ from bottomiun moments known to 3-loops using some judicious choices of moments from Ref. \cite{SNcb}. The errors on $m_b$ come respectively from the choice of the moments, $\alpha_s$, the data on the $\Upsilon$ family and the choice of the QCD continuum threshold. The ones due to the gluon condensates are negligible here.}
{\small
\begin{tabular}{ll}
&\\
\hline
%\hline
%\\
Mom&$m_b(m_b)$ [MeV] \\
%\\
\hline
%\hline
\\
$Q^2$=0:\\
$r_{2/3},~r_{2/4}$&$4160(4)(2)(3)(3)$\\
\\
$Q^2$=$4m_b^2:$ \\
$~r_{8/9},~r_{8/10}$&$4177(2)(3)(3)(6)$\\
\\
$Q^2$=$8m_b^2:$\\
$r_{13/14},~r_{13/15}$&$4183(2)(4)(2)(6)$\\
\\
\hline
Average &4173(4)\\
\hline
\end{tabular}
}
\label{tab:resb0}
\end{table}
}
\vspace*{-0.2cm}
\nin
%%%%%%%%%%%%%%%%%%%%%%%%%%%%%%%%%
%%%%%%%%%%%%%%%%%%%%%%%%%%%%%%%%
\section{ Conclusions}
\vspace*{-0.25cm}
%%%%%%%%%%%%%%%%%%%%%%%%%%%%%%%%%%%%%%%
 \nin
We summarize below the main results in this letter:\\
\b We have explicitly studied in Section \ref{sec:cont} the effect of the continuum model on the spectral function  and found that this effect is large for $Q^2=0$ low moments, which can only be evaded for moments ${\cal M}_{n\geq 3-4}(Q^2)$. This feature is naturally expected but raises the question on the errors induced by this model dependence in
the determinations of $\overline{m}_{c,b}$ from low-moments ${\cal M}_{n\leq 2}(0)$ used in the current literature.\\
\b We have extracted the value of $\la\alpha_s G^2\ra$ in Section \ref{sec:g2} and found the result in Eq. (\ref{eq:ag2}). This result confirms previous claims that the SVZ result underestimates the value of $\la\alpha_s G^2\ra$. We have not included in the analysis the most eventual short distance effect of the $D=2$ term advocated in \cite{CNZ,SNREV,SNZ1} which is dual to the higher order terms of the PT series \cite{SNZ2}. However, like in different explicit analysis of some other light quark channels, the effect of this term might also be small and can improve the agreement between the QSSR predictions with the data or with
some other determinations like lattice calculations. A future evaluation of this contribution is welcome but is beyond the scope of this paper. \\
\b We have re-extracted the gluon condensate $\la g^3 f_{abc} G^3\ra$ and obtained its value in terms of the instanton radius $\rho_c$ in Eq. (\ref{eq:rhoc}). This value agrees within the error with the one in \cite{SNcb} but is smaller than the estimate from the DIG approximation $\rho_c\simeq 5$ GeV$^{-1}$.\\
\b During the determinations of these condensates, our analysis prefers the value fac=0.5 of the $D=8$ $\la G^4\ra$ condensates which supports the modified factorisation proposed in \cite{BAGAN} using a $1/N$ approach.\\
\b We have re-estimated the $\overline{MS}$ running masses $\overline{m}_{c,b}$ to order $\alpha_s^3$ and including the $\la G^4\ra$ condensate contributions in the OPE. Optimal results from different moments lead to the final values in Eqs. (\ref{eq:mass_mc}) and  (\ref{eq:mass_mb}). These results confirm the recent results in \cite{SNcb} obtained from judicious choices of ratios of moments with small PT corrections and where the contributions of the $D=8$ $\la G^4\ra$ condensates have not been included. They also improve older results in \cite{SNmc} obtained at lower orders with larger errors. These results are also comparable with the ones in the existing literature using different methods \cite{IOFFE,KUHN2,MAIER,HOANG,KUHN1,BRAMB,LATTmc}.  

%\vfill\eject
 %%%%%%%%%%%%%%%%%%%%%%%%%%%%%%%
%%%%%%%%%%%%%%%

  \bibliography{mybib}
\end{document}